\documentclass[11pt,a4paper]{emulateapj} % use this line instead of pervious for two-column layouts
\usepackage{longtable}
\usepackage{enumerate}
\usepackage{float}
\usepackage{placeins}
\usepackage{amsmath}

\newcommand{\HI}{H\,{\sc i}~}
\newcommand{\HII}{H\,{\sc ii}~}

\begin{document}

\title{Resolving the electron temperature discrepancies in \HII Regions and Planetary Nebulae: $\kappa$--distributed electrons}

\shorttitle{$\kappa$--Distributions in \HII regions}
\shortauthors{Nicholls et al.}

\author{ David C. Nicholls\altaffilmark{1}, Michael A. Dopita\altaffilmark{1}\altaffilmark{2},  \&  Ralph S. Sutherland\altaffilmark{1}}
\email{david@mso.anu.edu.au}
\altaffiltext{1}{Research School of Astronomy and Astrophysics, Australian National University, Cotter Rd., Weston ACT 2611, Australia }
\altaffiltext{2}{Astronomy Department, King Abdulaziz University, P.O. Box 80203, Jeddah, Saudi Arabia}

\begin{abstract}
The measurement of electron temperatures and metallicities in \HII regions and Planetary Nebulae (PNe) has---for several decades---presented a problem:  results obtained using different techniques disagree.  What is worse, they disagree consistently.  There have been numerous attempts to explain these discrepancies, but none has provided a satisfactory solution to the problem.  In this paper, we explore the possibility that electrons in  \HII regions and PNe depart from a M-B equilibrium energy distribution.  We adopt  a ``$\kappa$--distribution'' for the electron energies. Such distributions are widely found in solar system plasmas, where they can be directly measured.  This simple assumption is able to explain the temperature and metallicity discrepancies in \HII regions and PNe arising from the different measurement techniques.  We find that the energy distribution does not need to depart dramatically from an equilibrium distribution. From an examination of data from \HII regions and PNe it appears that $\kappa  \gtrsim 10$ is sufficient to encompass nearly all objects.  We argue that the kappa-distribution offers an important new insight into the physics of gaseous nebulae, both in the Milky Way and elsewhere, and one that promises significantly more accurate estimates of temperature and metallicity in these regions.
\end{abstract}


\section{Introduction}
Over forty years ago, \citet{Vasyliunas68} measured the electron energy distributions in the Earth's magnetosphere using the OGO1 and OGO3 satellites and found that there was a significant high energy (`suprathermal') excess, when compared to an equilibrium M-B (M-B) distribution. The best fit to this high energy tail was a power law.  He introduced  the ``$\kappa$--distribution''---a generalized Lorentzian distribution---which provided a good description of the  electron energy distribution over the full range of energies.

Since then, $\kappa$--distributions have been the subject of considerable interest in solar system physics.  They have been found by direct measurement of electron energies by satellites and space probes in the outer heliosphere, the magnetospheres of all the gas-giant planets, Mercury and the moons Titan and Io, the Earth's magnetosphere, plasma sheet and magneto-sheath and in the Solar Wind (see references in \citet{Pierrard10}). Evidence is also emerging from IBEX observations that energetic neutral atoms in the interstellar medium, where it interacts with the helio-sheath, exhibit $\kappa$--distributions \citep{Livadiotis11a}. More than 5000 papers in many disciplines on the applications of $\kappa$--distributions had been published prior to 2011 \citep{Livadiotis11b}. In solar system plasmas, $\kappa$--distributions arise whenever the plasma is being continually pumped by an energy input of a non-thermal or supra-thermal nature, or by energy transport from elsewhere, so that the system cannot relax to a classical M-B distribution.

To date, despite the extensive adoption of $\kappa$--distributions in analysing solar system plasmas---where they are the rule rather than the exception---the possibility that electron energies are distributed in a non-thermal manner in the  photoionized plasma of \HII regions, Planetary Nebulae (PNe), or in the photoionized regions around active galaxies, does not appear to have been considered. Indeed the assumption that the electrons are in M-B equilibrium dates back over 70 years; see for example, \citet{Hebb40}. 

There are good reasons to question the basis on which temperatures and metallicities are measured in \HII regions and PNe: for example, there have been consistent discrepancies between direct electron temperature ($T_e$) estimates and those obtained using recombination lines.  None of the earlier attempts to solve the problem are fully satisfactory---for example, the ``$t^2$'' temperature fluctuation method, \citep{Peimbert67}---and some have been completely abandoned \citep{Stasinska04}.  

In this paper, we explore the implications of electron energies following a $\kappa$--distribution in these photoionized plasmas. We show that assuming a $\kappa$-- electron energy distribution is a simple and elegant way to resolve many of the difficulties, and should lead to more consistent temperature measurements and metallicity estimates in \HII regions and PNe.

The paper is organised as follows. In Section (2) we provide the key definitions and formulae for the $\kappa$--distribution. In Section (3) we show that there is a sound basis for  \HII regions and PNe having $\kappa$--distributed electrons. In Section (4) we examine the dependence of collisional excitation rates on excitation energy and the value of $\kappa$, and apply this to a detailed study of the particular case of the [\ion{O}{3}] $\lambda\lambda (4949+5007)/4363$ line ratio, the most important ratio used in the determination of electron temperatures in photoionized plasmas. In Section (5) we summarize the temperature discrepancy problem that has bedevilled temperature and metallicity measurements in gaseous nebulae for decades.  In Section (6) we examine the observational data from \HII regions and PNe to show how the assumption of a $\kappa$--distribution can explain the temperature discrepancies, and how the value of  $\kappa$ can be constrained by requiring that different methods of measuring the electron temperature yield the same answer.  Appendix A presents additional evidence for the presence of magnetic fields in \HII regions which can give rise to hot-tailed electron energy distributions.  In Appendix B, we examine the relationship between the ``$t^2$'' method for measuring the effect of temperature fluctuations and the $\kappa$--distribution.

\section{The $\kappa$--distribution}
\subsection{Properties and Definitions}
Initially, $\kappa$--distributions were used as an empirical fit to directly measured electron energies in solar system plasmas, and were criticized as  lacking a theoretical basis. More recently, the distribution has been shown to arise from entropic considerations. See, for example, \citet{Tsallis95, Treumann99, Leubner02}, and the comprehensive analysis by \citet{Livadiotis09}.  They explored the so-called $q$--nonextensive  entropy statistics in which the entropies of adjacent samples of plasma are not simply additive, and have shown that $\kappa$-- energy distributions arise naturally in such plasmas.  The requirement for this to occur is that there be long-range interactions between particles, in addition to the short-range Coulombic forces that give rise to M-B equilibration.  Although there is ongoing debate over whether the Tsallis statistical mechanics is the best generalisation of Boltzmann-Gibbs statistics, it provides a sound physical basis for the overtly successful use of the $\kappa$--distribution in plasma physics.

There are a number of slightly differing expressions for the energy distributions that can arise from $q$--nonextensive entropy, but the forms are generally similar. Here, we adopt the Vasyliunas form of the distribution as representative of the possible variants.  This is referred to by \citet{Livadiotis09} as a $\kappa$--distribution of the ``second kind''.  The successful use of the $\kappa$--distributions to describe physical phenomena in many disciplines, and especially in solar system physics, provides ample justification for exploring their use in \HII regions and PNe.

The $\kappa$--velocity distribution can be expressed (after Vasyliunas, 1968) as:
\begin{equation}\label{eq1}
n(v)\mathrm{d}v=\frac{4 N}{\sqrt \pi w_0^3} \frac{\Gamma(\kappa+1)}{\kappa^{3/2} \Gamma(\kappa-\frac{1}{2})} \frac{v^2}{(1 + v^2/(\kappa - \frac{3}{2})w_0^2)^{\kappa + 1}} dv\ ,
\end{equation}
where ${n(v)}$ is the number of electrons with speeds between ${v}$ and ${v+dv}$. The velocity ${w_0}$ is related to $w_{mp}$, the most probable speed  (i.e. the velocity value at the distribution peak) by:
\begin{equation}\label{eq2}
w_0 = w_{mp} \sqrt {\frac{\kappa}{ (\kappa - {\frac{3}{2}})}}\ .
\end{equation}
and related to the ``physical temperature'' of the system, $T_U$, as defined in \citet{Livadiotis09}:
\begin{equation}\label{eq3}
w_0 = \sqrt {\frac{2 k_BT_U}{m_e}}\ ,
\end{equation}
where ${m_e}$ is the electron mass and $k_B$ is the Boltzmann constant.  

$\kappa$ is a parameter that describes the extent to which the distribution departs from an equilibrium distribution.  In the Vasyliunas form, $\frac{3}{2} < \kappa \le \infty$. In the limit as $\kappa\to\infty$, the velocity distribution reverts to the standard M-B form,
\begin{equation}\label{eq4}
n(v) \mathrm{d}v=\frac{4 N}{\sqrt \pi w_0^3} v^2 \exp\left[-\frac{v^2}{w_0^2}\right] dv\ .
\end{equation}

The physical temperature, $T_U$, also referred to as the kinetic temperature, is a generalisation of the M-B equilibrium temperature, and is related to the energy density (system kinetic energy), $U$, which for a monatomic gas is the internal energy of the system:
\begin{equation}\label{eq5}
U = \frac{3}{2}k_BT_U\ .
\end{equation}

Figure 1 shows a family of $\kappa$-- velocity distributions with a M-B distribution.

\begin{figure}[htpb]
\includegraphics[width=\hsize]{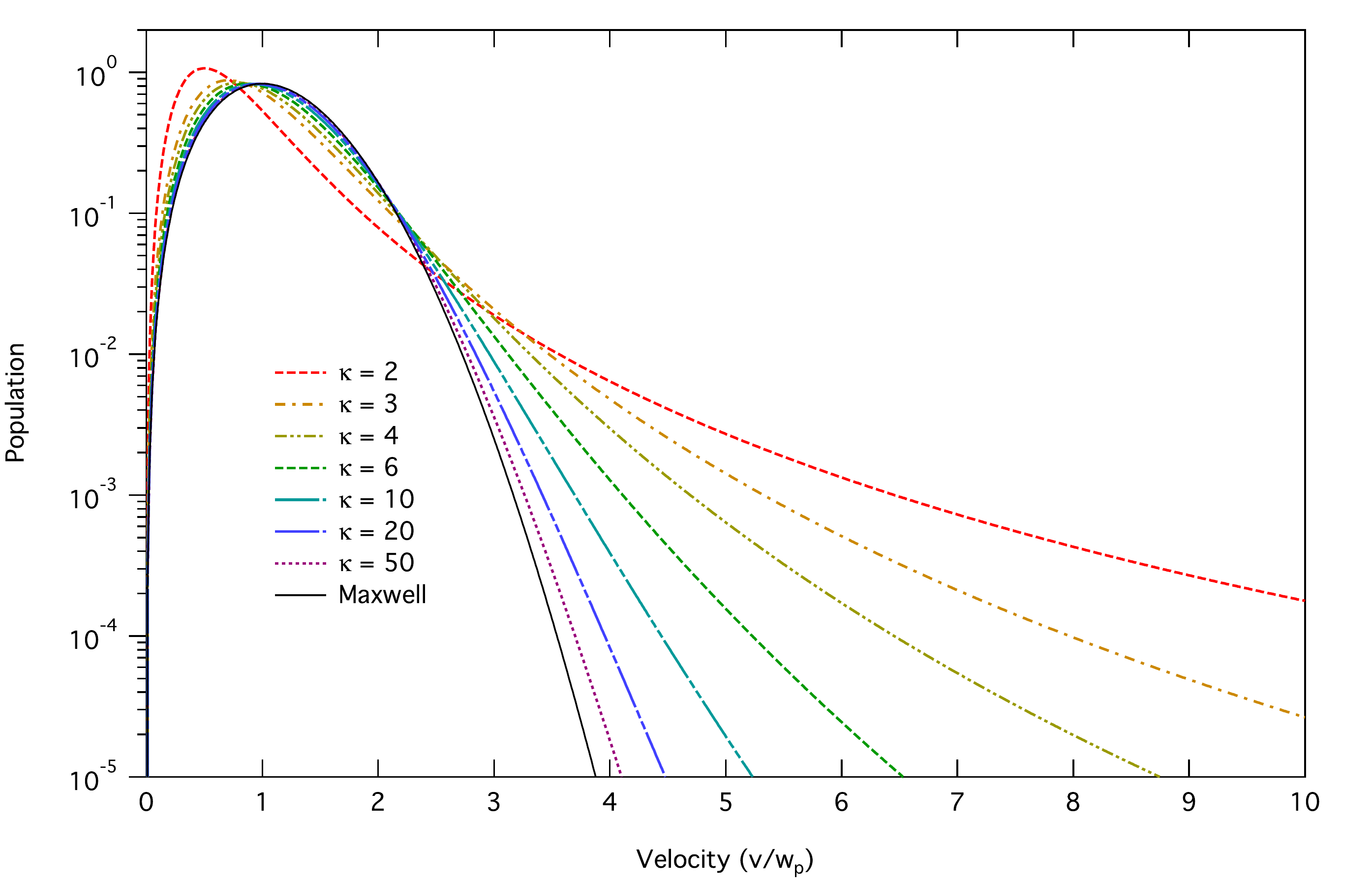}
\caption{$\kappa$-- velocity distributions (log scale) for $\kappa$ from 2 to 50, with Maxwell-Boltzmann distribution.}\label{fig_1}
\end{figure}

Expressed in energy terms, the $\kappa$--distribution becomes:
\begin{equation}\label{eq6}
\begin{aligned}
n(E) \mathrm{d}E=\frac{2 N}{\sqrt\pi} \frac{\Gamma(\kappa+1)}{(\kappa-\frac{3}{2})^{3/2} \Gamma(\kappa-\frac{1}{2})} \frac{1}{(k_BT_U)^{3/2}} \\
\times \frac{\sqrt E}{ (1 + E/((\kappa-\frac{3}{2}) k_BT_U))^{\kappa + 1}} dE\ .
\end{aligned}
\end{equation}

Again, the $\kappa$-- energy distribution tends in the limit as $\kappa\to\infty$ to the M-B,
\begin{equation}\label{eq7}
n(E) \mathrm{d}E=\frac{2 N}{\sqrt\pi} \frac{\sqrt E \ \exp\left[-E/k_BT\right]}{(k_BT)^{3/2}} dE\ .
\end{equation}

A family of normalized $\kappa$-- energy distributions is shown in Figure 2, together with a M-B equilibrium distribution. Note that the peak of the M-B energy distribution occurs at  $E = \frac{1}{2}k_B T_U$ whereas $\kappa$--distributions with the same internal energy peak at $E = \frac{1}{2}k_BT_U\ (2\kappa-3)/(2\kappa+1)$. 

\begin{figure}[htpb]
\includegraphics[width=\hsize]{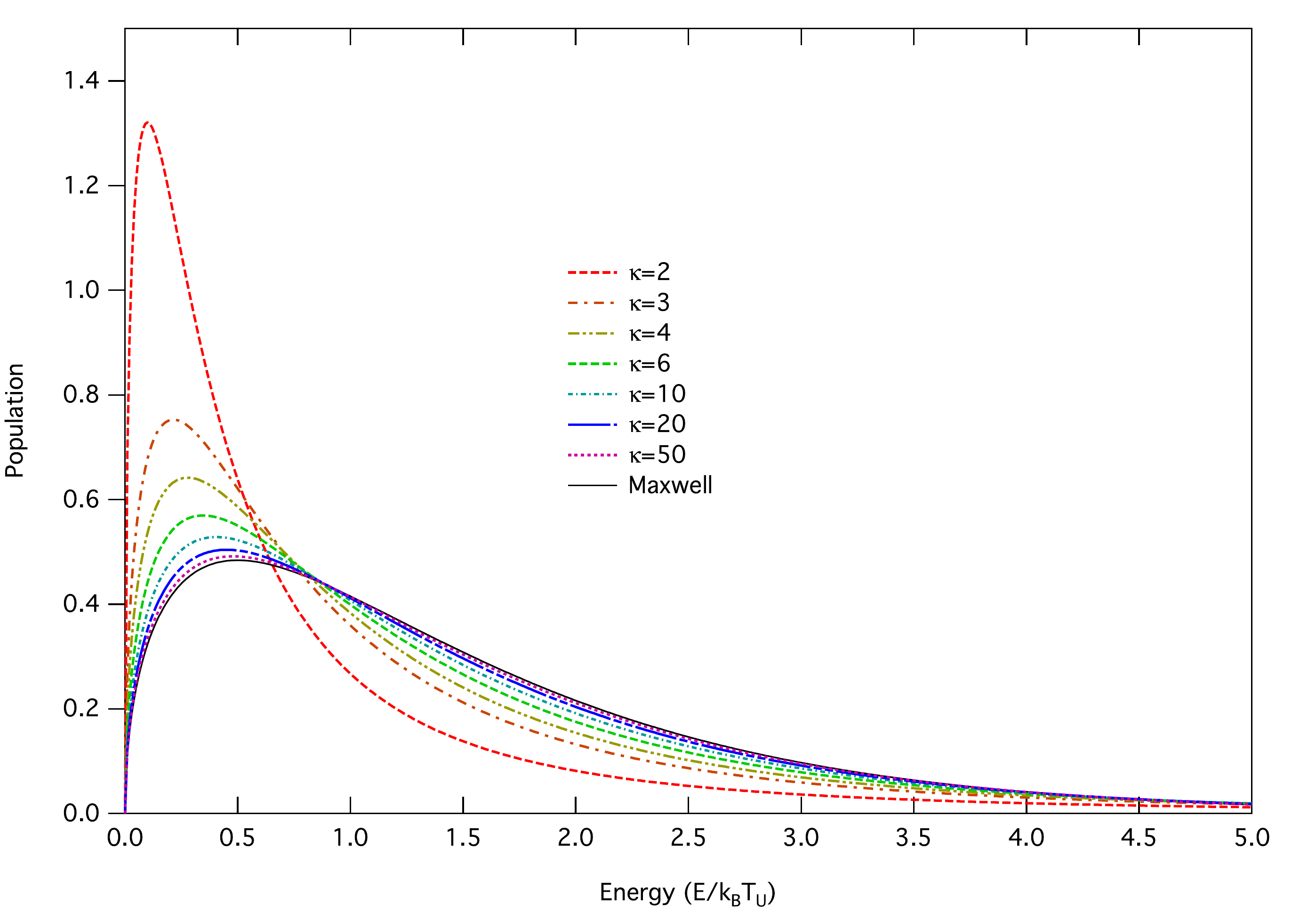}
\caption{$\kappa$-- energy distributions (linear scale) for $\kappa$ from 2 to 50, with Maxwell-Boltzmann distribution.}\label{fig_2}
\end{figure}
%\FloatBarrier % keep images together

The same distributions are shown in Figure 3 but plotted on a log scale.  The high energy power-law tail of the $\kappa$--distributions is clearly shown.

\begin{figure}[htpb]
\includegraphics[width=\hsize]{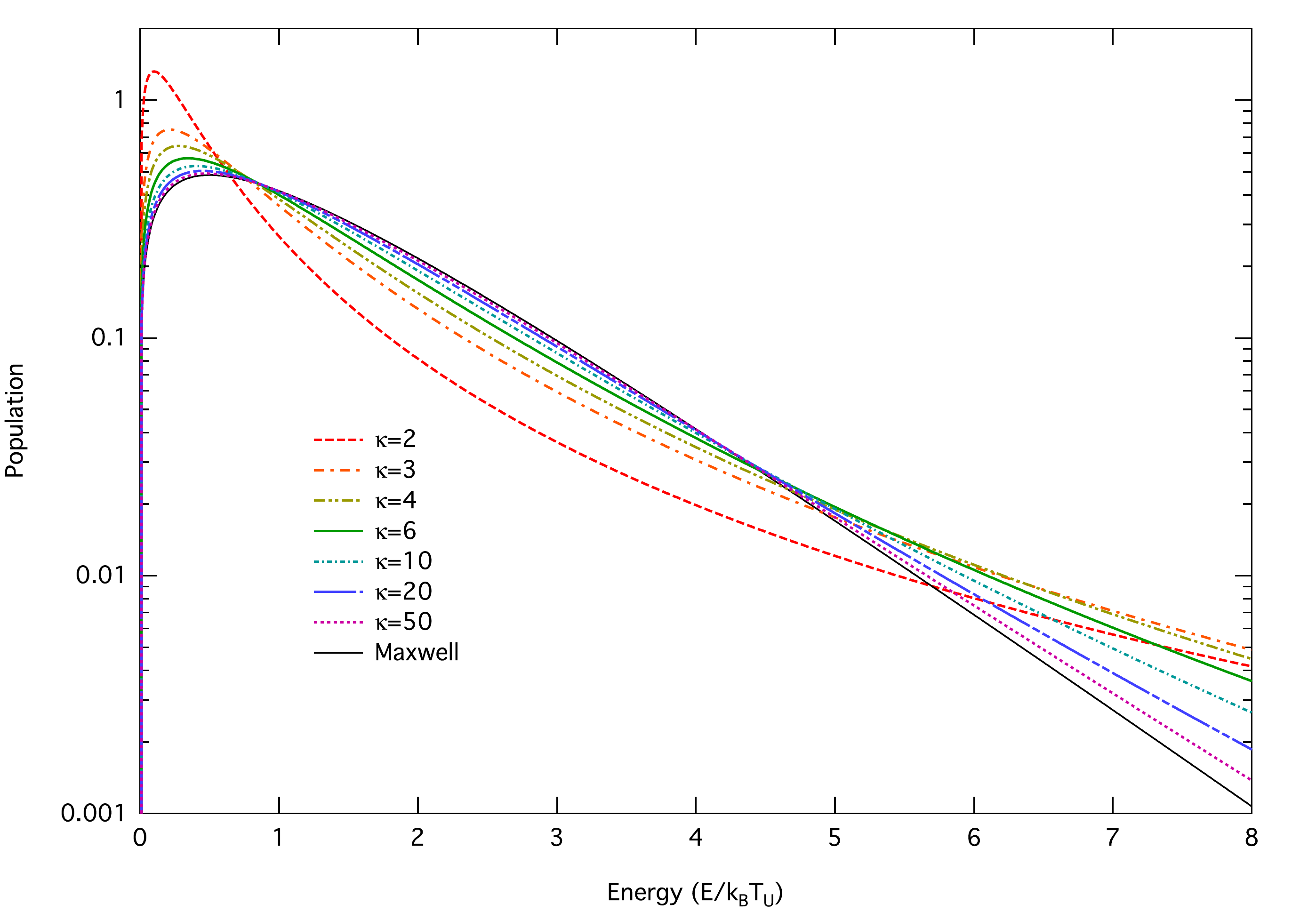}
\caption{$\kappa$-- energy distributions (log scale) for $\kappa$ from 2 to 50, with Maxwell-Boltzmann distribution.}\label{fig_3}
\end{figure}

\FloatBarrier % keep images together

Figures 1-3 illustrate the key characteristics of the $\kappa$--distribution: the peak of the distribution moves to lower energies; at intermediate energies there is a population deficit relative to the M-B distribution; and at higher energies the ``hot tail'' again provides a population excess over the M-B. The $\kappa$ distribution behaves as a M-B distribution at a lower temperature, but with a significant  high energy excess. 

This can be seen in Figure 4, where a M-B distribution is peak-fitted to a $\kappa$=2 distribution. The peak-fitted ``core'' M-B distribution is in fact at a lower physical temperature than the $\kappa$--distribution to which it is fitted.  The low energy ``core'' occurs because the equilibration timescale is a strong function of energy ($\propto \exp E^{3/2}$), allowing the low-energy electrons to approach an equilibrium distribution. The relationship between the physical temperature of the $\kappa$--distribution, $T_U$, and the equilibrium temperature of the ``core'' M-B, $T_{core}$, as implied by equation 3, is:

\begin{equation}\label{eq8}
T_{core}=T_U\left(\frac{\kappa-\frac{3}{2}}{\kappa}\right)\ .
\end{equation}

\begin{figure}[htpb]
\includegraphics[width=\hsize]{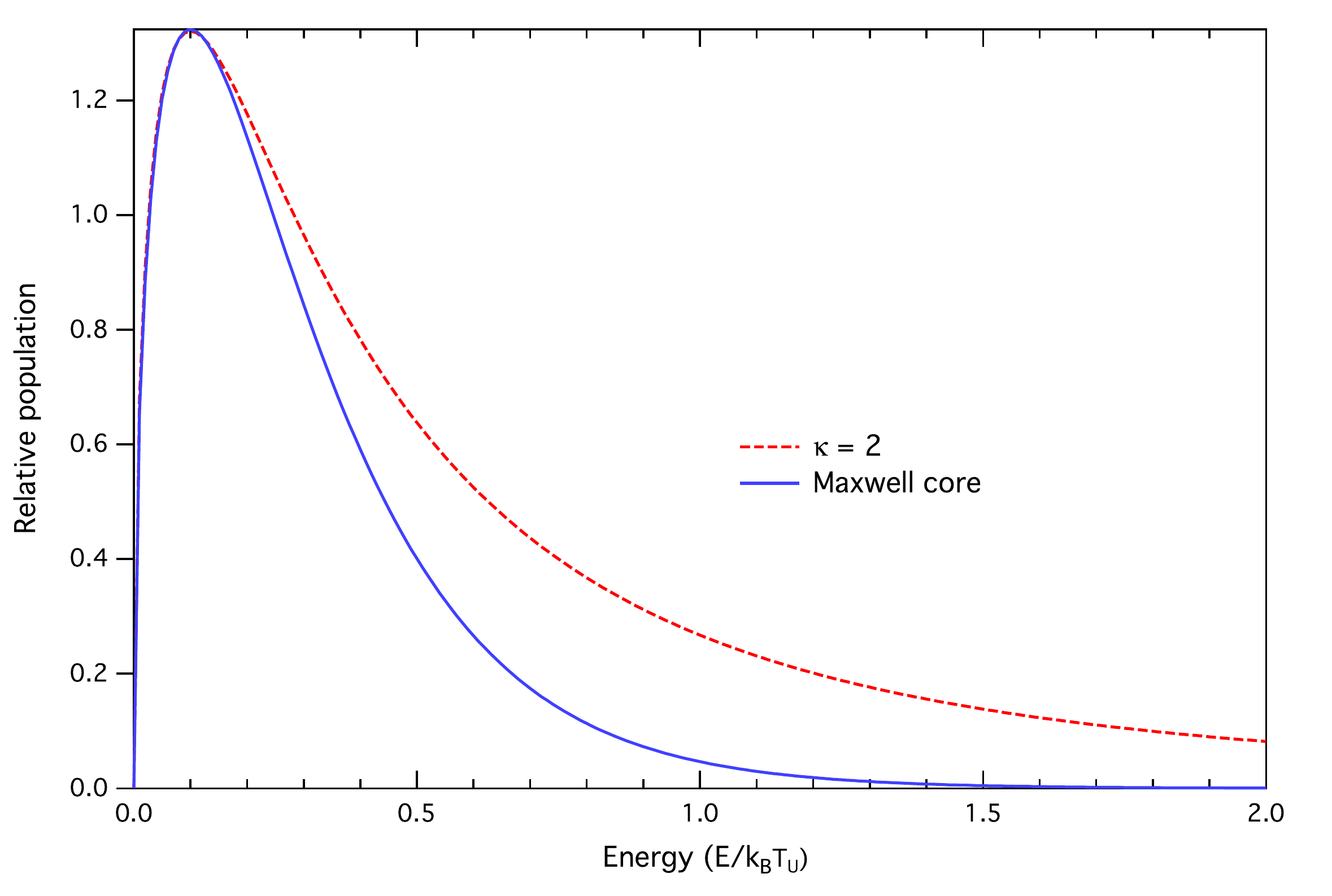}
\caption{$\kappa$ = 2 energy distribution with peak-fitted Maxwell-Boltzmann distribution core.}\label{fig_4}
\end{figure}

\FloatBarrier % keep images together

For physical processes that involve low energy electrons such as recombination line excitation, reactions ``see'' the cool M-B core distribution.   In other words, any physical property sensitive to the region of the electron energy distribution around or below the distribution peak will interact with an effective M-B electron energy distribution at a lower temperature than the M-B distribution with the same total internal energy as the $\kappa$--distribution.

Conversely, in processes that depend on the high energy end of the distribution, such as collisional line excitation,  the $\kappa$--distribution behaves like a M-B distribution at a higher temperature than one with the same total internal energy.  In other words, physical properties sensitive to energies above the peak will behave as if they were interacting with a hotter distribution. The $\kappa$--distribution thus exhibits a ``split personality", depending on the physical process involved. We show below that this behaviour resolves many of the discrepancies between different nebular temperature measurement techniques.

\section{What might give rise to $\kappa$--distributions in \HII regions?}

Given the plausiblity of $\kappa$--distributions in \HII regions, can we find mechanisms capable of maintaining the hot tails?  The answer is, clearly, yes. A high-energy  tail in the energy distribution will occur whenever the population of energetic electrons is being pumped in a timescale as short, or of the same order, as the energy re-distribution timescale of the electron population. Such effects may be long-range in nature, such as magnetic reconnection followed by the migration of high-energy electrons along field lines, and by the development of inertial Alfv\'en waves.    We explore additional evidence for the existence of magnetic fields in galactic \HII regions in Appendix A.

Pumping of high energy tails can also be more local in character such as local shocks (driven either by the collision of bulk flows or by supersonic turbulence), or, most simply, by the injection of high-energy electrons through the photoionization process itself. Normal photoionization produces supra-thermal electrons on a timescale similar to the recombination timescale.

Energetic electrons can also be generated by the photoionization of dust \citep{Dopita00}. Alternatively, X-ray ionization can produce highly energetic ($\sim$ keV) inner-shell (Auger process) electrons (e.g. \citet{Shull85,Aldrovandi85, Petrini97}, and references therein). Processes based on photoionization should become more effective where the source of the ionizing photons has a ``hard'' photon spectrum. Thus, the likelihood of the ionized plasma having a $\kappa$-- electron energy distribution would be high in the case of either photoionization by an Active Galactic Nucleus (AGN), or the case of PNe, in which the effective temperature of the exciting star could range up to $\sim 250,000$\degr K.

Thus, we have no shortage of possible energy injection mechanisms. The main consideration is whether feeding the energetic population can occur on a timescale which is short compared with the collisional re-distribution timescale, $\tau$. Because this timescale increases very rapidly with energy, $\tau \propto \exp(E^{3/2})$, we would expect there to be a threshold energy above which any non-thermal electrons have a long residence timescale. These can then feed continually down towards lower energies, maintaining a  $\kappa$-- electron energy distribution. It seems more likely therefore that \emph{all} photoionized plasmas will show departures from a M-B distribution to some degree. The key question is, is this departure important, and does it produce observable effects in the plasma diagnostics on which we have relied upon hitherto? Again, yes, as we show in Section (4) below.

\subsection{Are $\kappa$--distributions stable in their own right?}

In addition to the energy injection mechanisms capable of maintaining the excitation of  suprathermal distributions, several authors (\citet{Livadiotis11b} and references therein; \citet{Shizgal07, Treumann01}) have investigated the possibility that the $\kappa$--distribution may remain stable against thermalization longer than conventional thermalization considerations would suggest (e.g. \cite{Spitzer62}).  In particular, distributions with $2.5 \gtrsim \kappa > 1.5$ appear to have the capacity, through increasing entropy, of moving to values of lower $\kappa$ \citep{Livadiotis11b} i.e. away from (M-B) equilibrium.  While the physical application of this aspect of $\kappa$--distributions remains to be explored fully, it suggests that where $q$-nonextensive entropy conditions operate, the suprathermal energy distributions produced exist in ``stationary states'' where the behaviour is, at least in the short term, time-invariant \citep{Livadiotis10a}.  These states may have longer lifetimes than expected classically.  This is fully consistent with the numerous observations that in solar system plasmas, $\kappa$-- electron and proton energy distributions are the norm. It seems reasonable to expect that such conditions will also be present in \HII regions and PNe.

\section{The effect of $\kappa$--distributions on collisional excitation in \HII regions}

\subsection{Collisional excitation rates}
Consider the collisional excitation of an atomic species from energy level $1$ to energy level $2$.  The rate of collisional population of the upper energy level per unit volume is given in terms of the collision cross-section, $\sigma_{12}(E)$, by:

\begin{equation}\label{eq9}
R_{12}=n_e N_1 \int\limits_{E_{12}}^\infty \sigma_{12}(E) \sqrt E f(E)\ dE\ .
\end{equation}

We separate out the strong energy dependence of the collision cross-section $\sigma_{12}(E)$, by expressing it in terms of the collision strength, $\Omega_{12}$ and the energy, $E$:

\begin{equation}\label{eq10}
\sigma_{12}(E)=\left(\frac{h^2}{8\pi m_e E}\right)\frac{\Omega_{12}}{g_1}\ ,
\end{equation}
where $h$ is the Planck Constant, $m_e$ is the electron mass, and $g_1$ is the statistical weight of the lower energy state.

The collisional population rate now becomes:
\begin{equation}\label{eq11}
R_{12}=n_e N_1  \frac{h^2}{8 \pi m_e g_1}\int\limits_{E_{12}}^\infty\frac{\Omega_{12}}{\sqrt E}\ f(E)\ dE\ ,
\end{equation}
where the appropriate form of the distribution is substituted, giving the well-known collisional population rate formula for a M-B distribution:
\begin{equation}\label{eq12}
\begin{aligned}
R_{12}(\mathrm{M-B})=n_e N_1  \frac{h^2 }{4 \pi^{3/2} m_e g_1} \left(k_B T_U \right)^{-3/2} \\
\times  \int\limits_{E_{12}}^\infty \Omega_{12} \ exp\left[-\frac{E}{k_BT_U}\right] dE\ .
 \end{aligned}
\end{equation}
For a $\kappa$--distribution, the corresponding rate is:
\begin{equation}\label{eq13}
\begin{aligned}
R_{12}(\kappa)=n_e N_1  \frac{h^2}{4 \pi^{3/2} m_e g_1} \frac{\Gamma(\kappa+1)}{(\kappa-\frac{3}{2})^{3/2}\Gamma(\kappa-\frac{1}{2})} \left({k_BT_U}\right)^{-3/2} \\
\times \int\limits_{E_{12}}^\infty \frac{\Omega_{12}}{(1 + E/[(\kappa-\frac{3}{2}) k_BT_U)]^{\kappa + 1}}\ dE\ .\  \  \ 
\end{aligned}
\end{equation}

Adopting the approximation that $\Omega_{12}$ is independent of energy, the integral parts of the above equations (after taking a $k_BT_U$ factor outside the integrals) reduce to:
\begin{equation}\label{eq14}
\exp\left[-\frac{E_{12}}{k_BT_U}\right]
\end{equation}
and
\begin{equation}\label{eq15}
\left (1-\frac{3}{2\kappa}\right ) \left( 1+\dfrac{E_{12}}{(\kappa-\frac{3}{2})k_BT_U} \right) ^{-\kappa},
\end{equation}
respectively. Note that in the limit as $\kappa \rightarrow \infty$, equation $\eqref{eq15}$ transforms into equation $\eqref{eq14}$, as it should.

It is useful to compare the relative rates of population to an upper state for M-B and $\kappa$--distributions.  Taking equations \eqref{eq12} and \eqref{eq13} and assuming constant $\Omega$s, we can derive an analytical equation that expresses the ratio of the population rates:

\begin{equation}\label{eq16}
\begin{aligned}
\dfrac{R_{12}(\kappa)}{R_{12}(\mathrm{M-B})}=  \frac{\Gamma(\kappa+1)}{(\kappa-\frac{3}{2})^{3/2}\Gamma(\kappa-\frac{1}{2})}\left(1-\frac{3}{2\kappa} \right) \\
\times \exp \left[\frac{E_{12}}{k_BT_U}\right] \left( 1+\dfrac{E_{12}}{(\kappa-\frac{3}{2})k_BT_U)} \right) ^{- \kappa}\ .
\end{aligned}
\end{equation}

Figure 5 shows the effect of equation \eqref{eq16} in the enhancement of the collisional excitation rate for $\kappa$: 2 $\to$ 100 compared to the M-B rate for an electron distribution having the same internal energy. Very similar curves were obtained by \citet{Owoki83} in the context of collisional ionization rates in the solar corona.  Data for Figure 5 for an extended energy ratio range are given in Table 1.
\begin{figure}[htpb]
\includegraphics[width=\hsize]{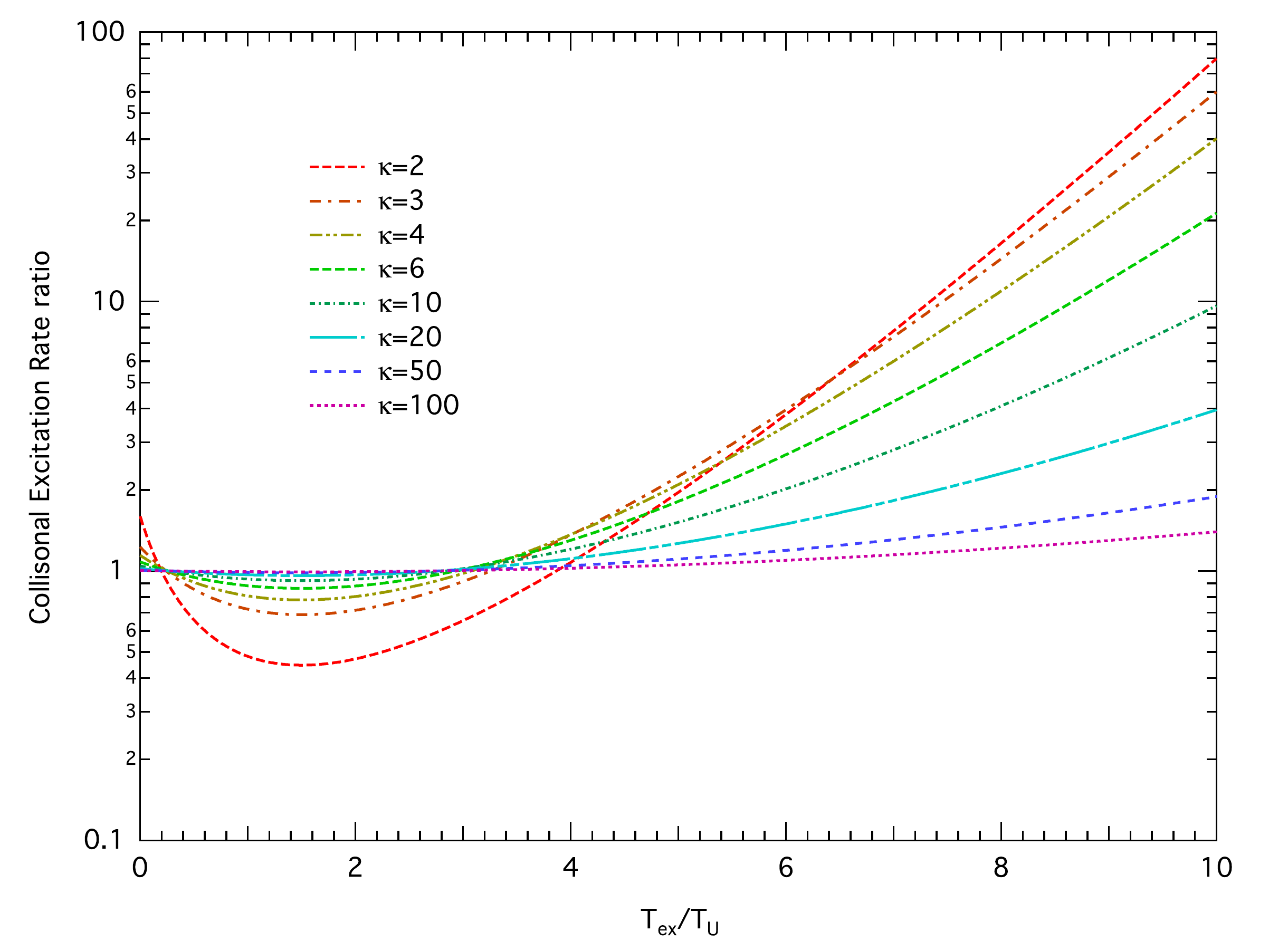}
\caption{The collisional excitation enhancement ratio over a Maxwell-Boltzmann distribution for different $\kappa$--distributions plotted as a function of the excitation threshold energy (expressed as an equivalent temperature) over the kinetic temperature $T_U$, equation \eqref{eq16}. For the $\kappa$--distributions very large enhancements in the collisional excitation rate are possible for high values of $T_{\rm ex}/T_U$.  (Extended data in Table 1)}\label{fig_5}
\end{figure}

The implication is that collisionally excited UV lines---with higher excitation energies--- should show strong enhancements compared to standard theory, while lines in the optical and IR would be relatively little changed, unless the region has a low kinetic temperature. Collisionally excited lines of non-ionized species such as [\ion{O}{1}] $\lambda\lambda 6300,63$ or [\ion{N}{1}] $\lambda\lambda 5198,5200$ will be differently affected, since the collision strength has a strong energy dependence, increasing above threshold. For these ions the degree of enhancement in the collisional excitation rate compared with the M-B case will be much larger than shown in Figure 6. This could explain a long-standing problem in \HII region models, which tend to systematically underestimate the strength of these lines, compared to observations.  New data on the collision strengths of OI and NI, in the context of the $\kappa$--distribution, have implications for using the lines of these neutral species as low-temperature diagnostics in partially ionized regions.  We will explore this in a future paper.

\subsection{Effect of $\kappa$ on [\ion{O}{3}] electron temperatures}

The direct method of estimating electron temperature in \HII regions relies on measuring the flux ratios of different excited states from the same atomic species.  Most often used are collisional excitation to the  $^1D_2$ and $^1S_0$ states of [\ion{O}{3}] , which give rise to the forbidden nebular lines at 5007\AA \ and 4959\AA, and the ``auroral''  line at 4363\AA. Lines of [\ion{O}{2}], [\ion{N}{2}], [\ion{S}{2}] , [\ion{Ne}{3}], [\ion{Ar}{3}] and [\ion{S}{3}]  are also used, when the relevant lines can be observed. The threshold excitation energies for the upper states in these species are different, so the degree of enhancement in a $\kappa$--distribution of electrons will differ from one ion to the next. This fact provides a possible resolution of the discrepancies encountered when measuring temperatures using these different atomic species.

\begin{figure}[htpb]
\includegraphics[width=\hsize]{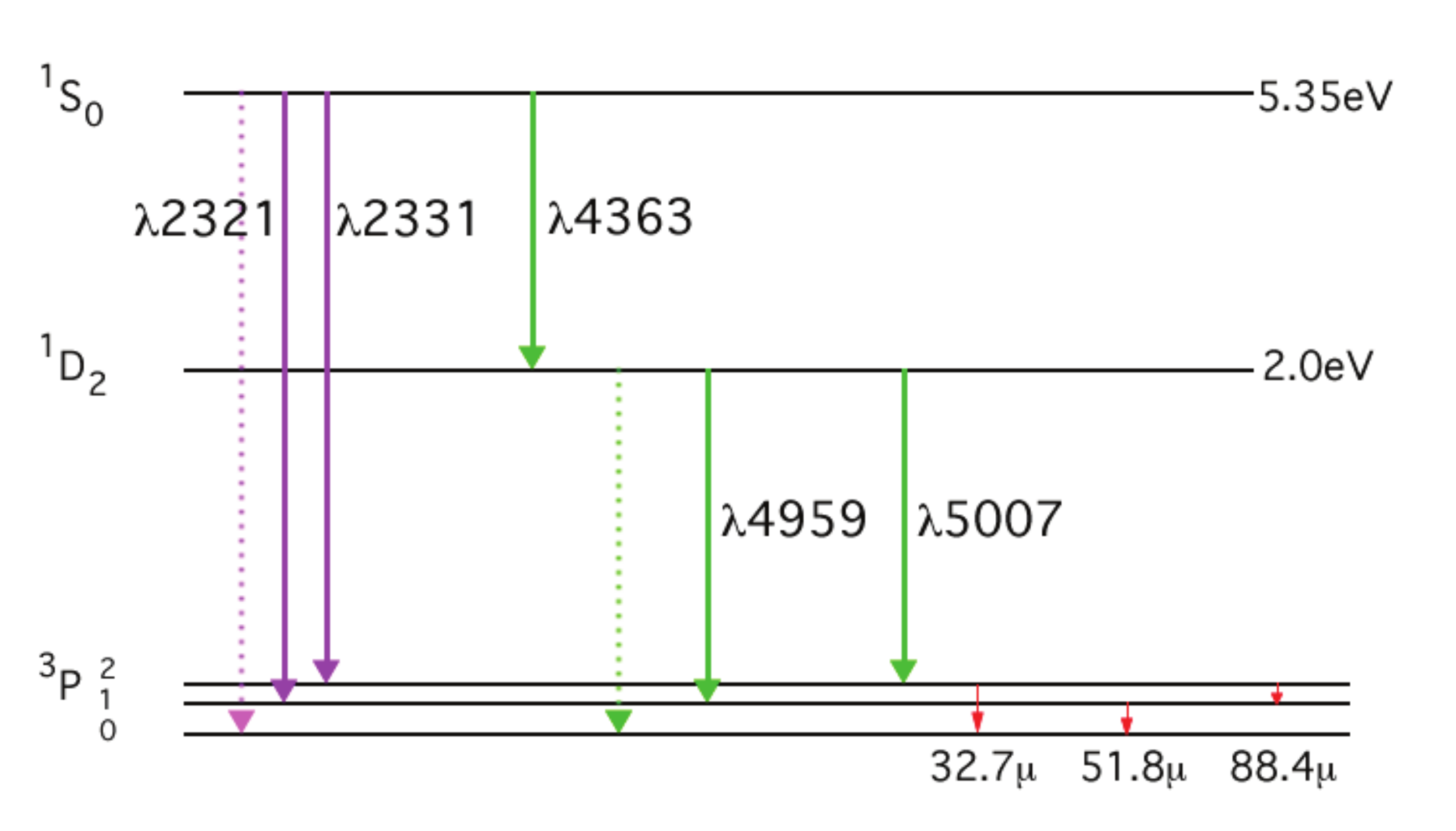}
\caption{The lower energy levels and forbidden transitions for the \ion{O}{3} ion.}\label{fig_6}
\end{figure}
\FloatBarrier % keep images together

As a specific and important example, we consider here the excitation of the \ion{O}{3} ion.  For reference, the configuration of the lower states which give rise to the forbidden lines used in temperature determinations is illustrated in Figure 6. The threshold excitation temperatures of the excited states are 29,130\degr K ($^1D_2$) and 62,094\degr K ($^1S_0$). For a $\kappa$--distribution at temperature $T_U=10,000$\degr K from Figure 5 the excitation rate to the lower state will be little changed, while the collision excitation rate to the upper state will be strongly enhanced, leading to an overestimate of the true electron temperature computed by formulae such as given by  \citet{Osterbrock05}. We now proceed to quantify this remark.

For an equilibrium electron energy distribution, the relative population rates can be calculated using equation \eqref{eq11} for the upper and lower excited states.  If one assumes for simplicity that the $\Omega$ values are energy-independent, the ratio of the two population rates in the M-B case is simply:
\begin{equation}\label{eq17}
\frac{R_{13}}{R_{12}}=\frac{\Omega_{13}}{\Omega_{12}} \exp\left[-\frac{E_{23}}{k_BT_U}\right]\ .
\end{equation}
In a more accurate analysis, the collision strength $\Omega$ is integrated over the energy range $E_{12}$ to $\infty$ as per equation \eqref{eq11}, to give the effective (temperature averaged) collision strength, $\Upsilon$.

Allowing for other transitions from the  $^1S_0$ state via the branching ratio and correcting for the energies of the two lower state transition photons, the ratio of the fluxes of the ($\lambda$5007 +  $\lambda$4959) to  $\lambda$4363 give a direct measure of the population rates to the two excited states, and therefore of the electron temperature of the energy distribution. Following \citet{Osterbrock05} we can derive a simple equation for the electron temperature, $T_e$, in terms of the line fluxes, for low density plasmas:

\begin{equation}\label{eq18}
\frac{j5007+j4959}{j4363}=7.90 \exp\left[\frac{32900}{T_e}\right]\ .
\end{equation}

This formula is, of course, an approximation, as it assumes the values $\Omega_{12}$ and $\Omega_{13}$ (and therefore of $\Upsilon_{12}$ and $\Upsilon_{13}$) do not depend on temperature.  A more accurate iterative formula was given by \citet{Izotov06}.  Values of the effective collision strength $\Upsilon$ assuming constant $\Omega$ differ slightly from those computed numerically, taking into account the detailed collisional cross section resonances.  To identify what effect this would have on determining the electron temperature, we compared electron temperatures derived using constant $\Omega$, the Izotov iterated formula, and the values computed using the detailed collision strengths from \citet{Lennon94} and \citet{Aggarwal93}.  

Figure 7 shows the relationship between the flux ratio and the electron temperature, using the Osterbrock equation, the Izotov iteration and computed numerically from the \citet{Lennon94} and \citet{Aggarwal93} $\Omega$ data. (The values in most common use for $\Upsilon_{12}$ are those published by \citet{Lennon94}, available online via TIPbase \citep{Hummer93}; or the data from \citet{Aggarwal99}, but the latter are only available in abbreviated tabular form.)

\begin{figure}[htpb]
\includegraphics[width=\hsize]{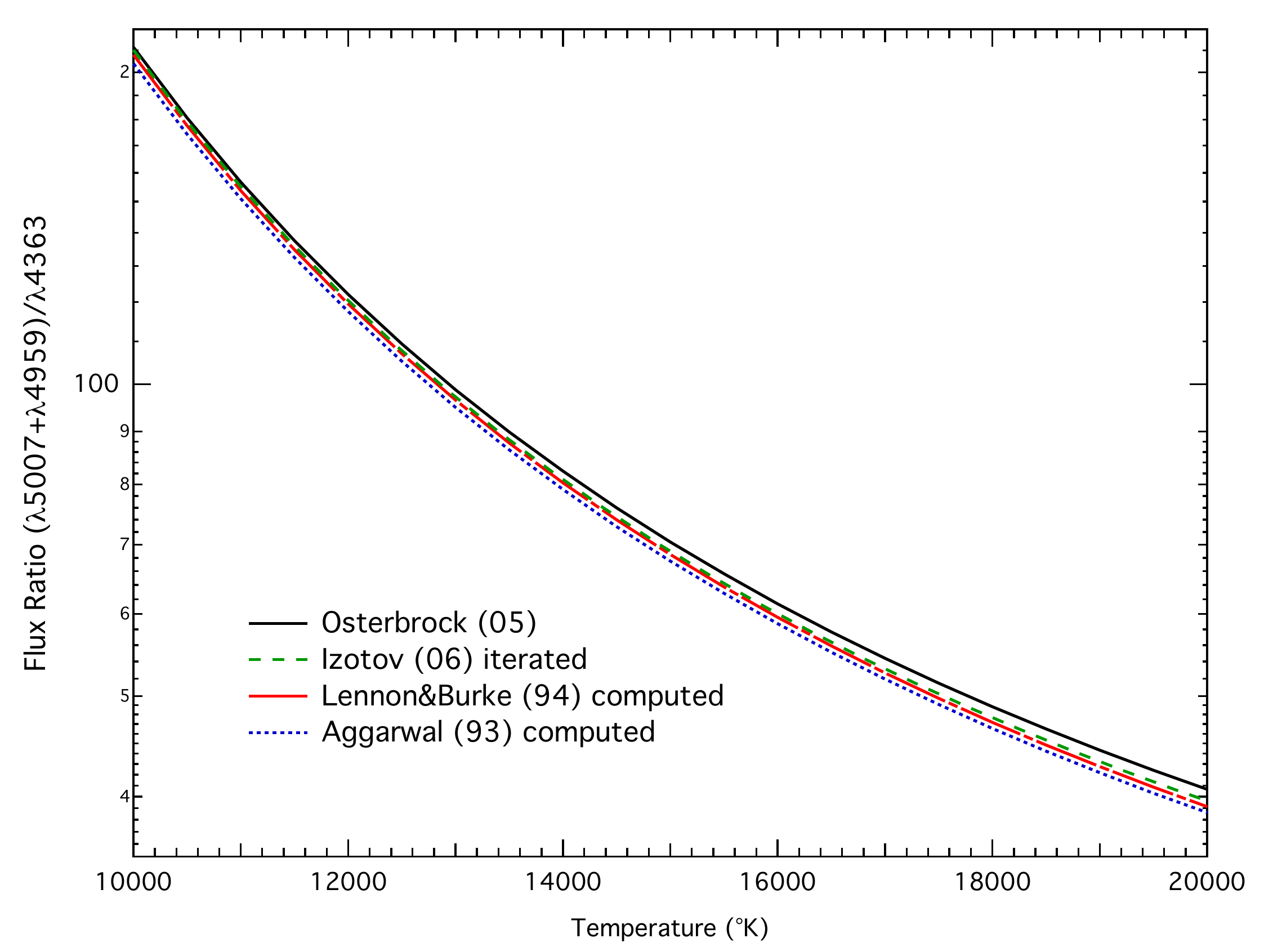}
\caption{[\ion{O}{3}] line flux ratio vs. temperature for M-B distributed electrons computed from different methods and by different authors}\label{fig_7}
\end{figure}
\FloatBarrier % keep images together

For M-B energy distributions, the use of a constant $\Omega$ gives a slightly higher $T_e$ than is obtained using the \citet{Lennon94} data, and higher still compared to the data from \citet{Aggarwal93}.  At an equilibrium temperature of 20,000\degr K, the difference between the Izotov value and the L\&B data is 130\degr K and for the Aggarwal data, 285\degr K.

These differences are of minor import. An altogether different result occurs when we use a $\kappa$--distribution instead of the M-B to calculate the line ratio vs $T_U$ graph.  The results are shown in Figure 8.  For $\kappa$=100, 50, 20, and 10 the kinetic temperature differences from the M-B equilibrium value at $T_U$=20,000\degr K are 180\degr K, 385\degr K, 980\degr K and 2,100 \degr K.   Data for the extended temperature range 5,000 to 20,000\degr K are given in Table 2.  The data are calculated using the detailed collision strengths (from \citet{Lennon94}), but assuming constant $\Omega$s makes only a small difference for $\kappa >$ 6.

As can be seen in Figures 2 and 3, these values of $\kappa$ are visually relatively minor deviations from the M-B distribution, but they have a considerable effect.  This implies that even $\kappa$--distributions which diverge slightly from equilibrium can have a significant effect on electron temperatures measured using the [\ion{O}{3}] lines.  The same result is true for other collisionally excited lines, [\ion{O}{2}], [\ion{S}{2}] and [\ion{N}{2}], but to differing extents, owing to the different collisional excitation energy thresholds for the upper and lower states.  The differences in apparent electron temperatures calculated using different collisionally excited species allow us to obtain an estimate of the effective value of $\kappa$ for remote \ion{H}{2} regions, as we show in Section (6).

\begin{figure}[htpb]
\includegraphics[width=\hsize]{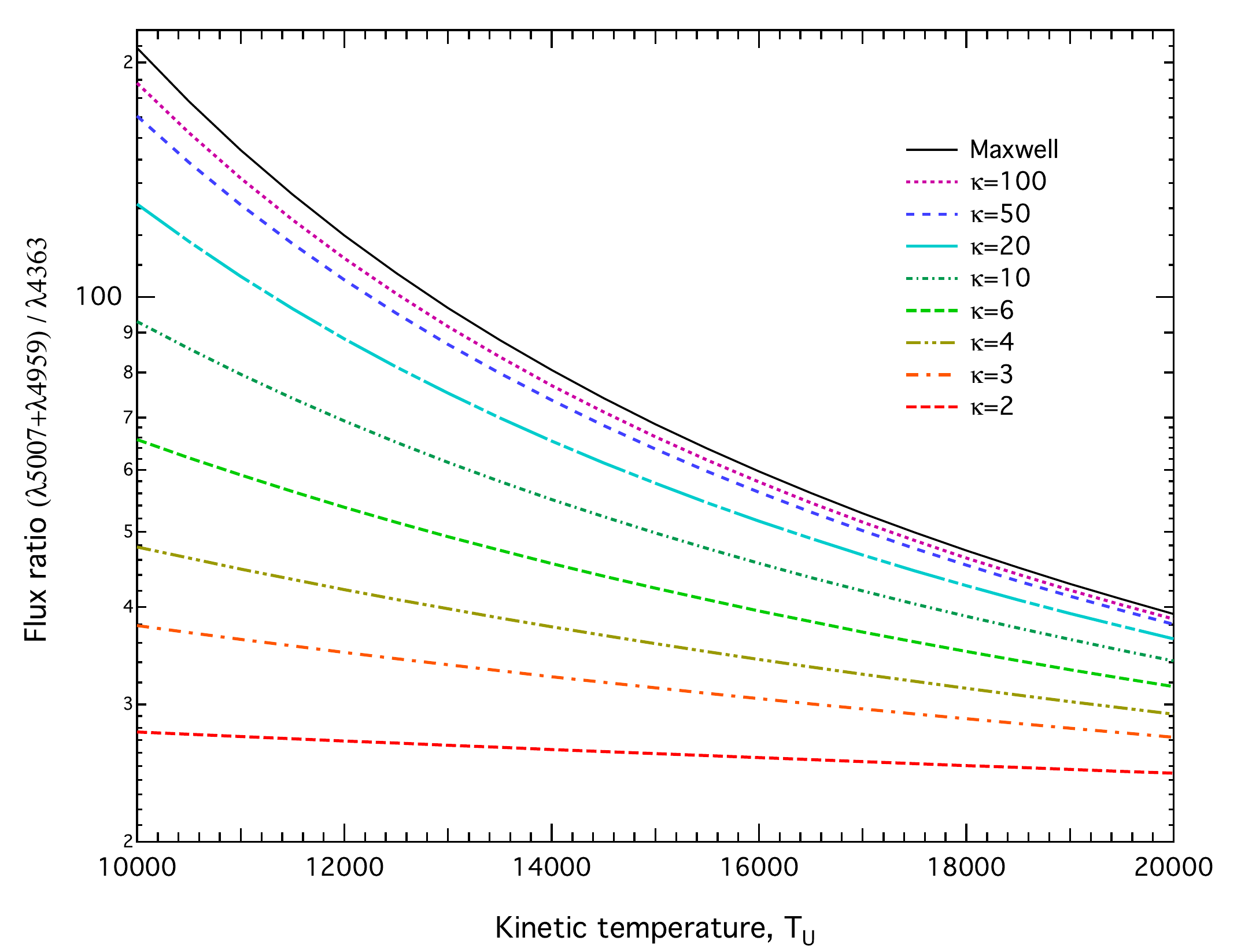}
\caption{[\ion{O}{3}] line flux ratio vs kinetic temperature for the range $\kappa=2 \to100$, compared to the computed \citet{Lennon94} equilibrium data, labeled Maxwell in the figure.  (Extended data in Table 2)}\label{fig_8}
\end{figure}
\FloatBarrier % keep images together
 
 \section{The abundance discrepancy problem}
 
Discrepancies between measurements made using various temperature-sensitive ion line ratios have been observed in both \HII regions and PNe. Recombination temperatures (derived from the Balmer and Paschen jumps) are systematically lower than temperatures derived from collisionally excited line ratios. Systematic differences in temperature are also observed between collisionally excited line ratios of different ionic species  \citep{G-R07,Izotov06}. This results in systematically different chemical abundances derived by optical recombination lines (ORLs) and collisionally excited lines (CELs), a problem dating back to \citet{Wyse42}, and first discussed in detail by \citet{T-P77}. There are also systematic differences between abundances determined using either direct measurements of ionic temperatures, correcting observed ionic abundances for unseen ionization stages (the $T_e$ method) or using the intensities of strong emission lines relative to the Hydrogen recombination lines (the strong line method). Collectively these are  known as the abundance discrepancy problem \citep{G-R06,G-R07}.

The confusion surrounding this issue has been well summarised by \citet{Stasinska04}:
\begin{quotation}``It has been known for several decades that optical recombination lines [ORL] in PNe and \HII regions indicate higher abundances than collisionally excited lines [CEL]. ... ... The ORL abundances are higher than CEL abundances by factors of about 2 for most PNe, discrepancies over a factor 5 are found in about 5\% of the PNe and can reach factors as large as 20. For a given nebula, the discrepancies for the individual elements C,N,O,Ne are found to be approximately of the same magnitude.

The explanations most often invoked are: i) temperature fluctuations, ii) incorrect atomic data, iii) fluorescent excitation, iv) upward bias in the measurement of weak line intensities, v) blending with other lines, vi) abundance inhomogeneities.  None of them is completely satisfactory, some are now definitely abandoned."
\end{quotation}

The possibility that the abundance discrepancy problem could be generated by the techniques used to measure abundances was recently investigated by \citet{Lopez12} using a grid of theoretical models to eliminate any systematics of observational errors. Although systematic errors were found in a number of empirical strong line techniques, the classical electron temperature + ionization correction factor technique works surprisingly well. Clearly the technique itself is not at fault. 

Perhaps the most successful attempt to account for the abundance discrepancy problem has been to postulate the existence of small-scale temperature fluctuations as first proposed by \citet{PC69}, known as the ``$t^2$'' method, and applied widely since; see \emph{e.g.} \citet{PeiM-Bert03} and references therein. To some extent electron temperature fluctuations (if real) act in a similar way to a  $\kappa$--distributed electron population, in that both weight CELs towards higher temperature. However, we are missing an obvious physical explanation of why micro-fluctuations in temperature should exist and persist in the first place.

The $\kappa$ approach has major advantages over the $t^2$ method:  it is simple; it is consistent over many objects and atomic species; it arises directly from Tsallis q-nonextensive statistics; the physics of ionized plasmas in \HII regions and PNe provides several mechanisms capable of generating the suprathermal distribution tails; it has been shown to describe accurately numerous directly measured electron energy distributions in the solar system; and it explains behaviour over an energy range of at least 3 orders of magnitude, from low energy recombination line electrons to the high ionization energies in the solar corona. The relationship between the $t^2$ approach and the $\kappa$--distribution is explored further in Appendix B.

\section{The Determination of $\kappa$}
\subsection{ \HII regions}
As noted earlier (equation 8), a $\kappa$--distribution with kinetic temperature $T_{\rm U}$ can be characterised at low electron energies below the peak of the distribution by a ``core'' Boltzmann distribution with an effective temperature $T_{core}=(\kappa-3/2)T_{\rm U}/\kappa$. Since $T_{core}$  is systematically lower than  $T_{\rm U}$, and as it is the energy distribution of these low-energy electrons that determines the recombination temperature ($T_{\rm rec} \equiv T_{core}$), it is clear that in a $\kappa$--distribution the recombination rate is systematically enhanced. These same low-energy electrons also determine the size of the Balmer and Paschen discontinuities of Hydrogen, so that the measured Balmer and Paschen break temperatures will reflect  $T_{B}$, rather than $T_{\rm U}$.

At the same time, lines with excitation temperatures comparable to $T_{\rm U}$ may be either mildly enhanced or suppressed in a $\kappa$--distribution, but lines with excitation temperatures well above $T_{\rm U}$ are strongly enhanced by the power-law tail of high energy electrons present in the $\kappa$--distribution. This has the effect of enhancing the apparent electron temperature inferred using well-known temperature-sensitive line ratios such as [\ion{O}{3}] 4363\AA/5007,4959\AA, [\ion{S}{2}] 4069,76\AA/6717,31\AA,  [\ion{N}{2}] 5755\AA/6548,84\AA\  or  [\ion{O}{2}] 7318,24\AA/3726,29\AA. The degree of enhancement of the inferred temperature, compared to $T_{\rm U}$, is strongly dependent on $T_{\rm ex}$, the excitation temperature of the upper levels involved in these transitions. Those ions in which $T_{\rm ex}$ is much greater than $T_{\rm U}$ will have their inferred collisional excitation temperature,  $T_{\rm CEL}$, very strongly  enhanced over $T_{\rm U}$, while those for which  $T_{\rm ex}$ is comparable to  $T_{\rm U}$ will show little change ($T_{\rm CEL} \sim T_{\rm U}$).

These properties of the $\kappa$--distribution can be used as the basis for the determination of $\kappa$ from spectrophotometric observations of  both \ion{H}{2} regions and Planetary Nebulae. 

We can test the $\kappa$--distribution hypothesis using the excellent echelle spectrophotometry which has been gathered by a number of authors in recent years. For Galactic \ion{H}{2} regions we have data in M 42 \citep{E04}, NGC 3576 \citep{G-R04}, S311 \citep{G-R05}, M20 \& NGC 3603 \citep{G-R06}, M 8 \& M 17 \citep{G-R07}. For the extragalactic  \ion{H}{2} regions we have data for 30 Dor  \citep{PeiM-Bert03}, NGC5253 \citep{L-S07}, NGC 595, NGC 604, VS 24, VS 44, NGC 2365 and K 932 \citep{E09}.

One of the strongest tests of the validity of the $\kappa$--distribution is the comparison of the  [\ion{O}{2}] 7318,24\AA/3726,29\AA\ and [\ion{S}{2}] 4069,76\AA/6717,31\AA\ temperatures. These ions have very similar ionization potentials and are therefore distributed in a very similar way in the nebula. The photoionization models used by \citep{Lopez12} have line emission weighted temperatures in the \ion{O}{2} and \ion{S}{2} zones which differ from each other by less than 400\degr K over the abundance range $0.3-3.0$ times solar. However the excitation temperatures of the lines are quite different. For the  [\ion{S}{2}] 4069,76\AA\ lines, $T_{\rm ex} = 35,320$\degr K, and for the  [\ion{S}{2}] 6717,31\AA\ $T_{\rm ex} = 21,390$\degr K. In the case of [\ion{O}{2}] 7318,24\AA, $T_{\rm ex} = 58,220$\degr K while for the 3726,29\AA\ lines, $T_{\rm ex} = 38,590$\degr K. Thus the  [\ion{O}{2}] lines are much more sensitive to the high-energy tail of the $\kappa$--distribution than are the  [\ion{S}{2}] lines.  On this basis, one would expect higher temperatures to be derived from the [O II] line ratios than from the [S II] line ratios, in the presence of a $\kappa$--distribution.
\begin{figure*}[ht] % Figure 9 full width
\includegraphics[width=\hsize]{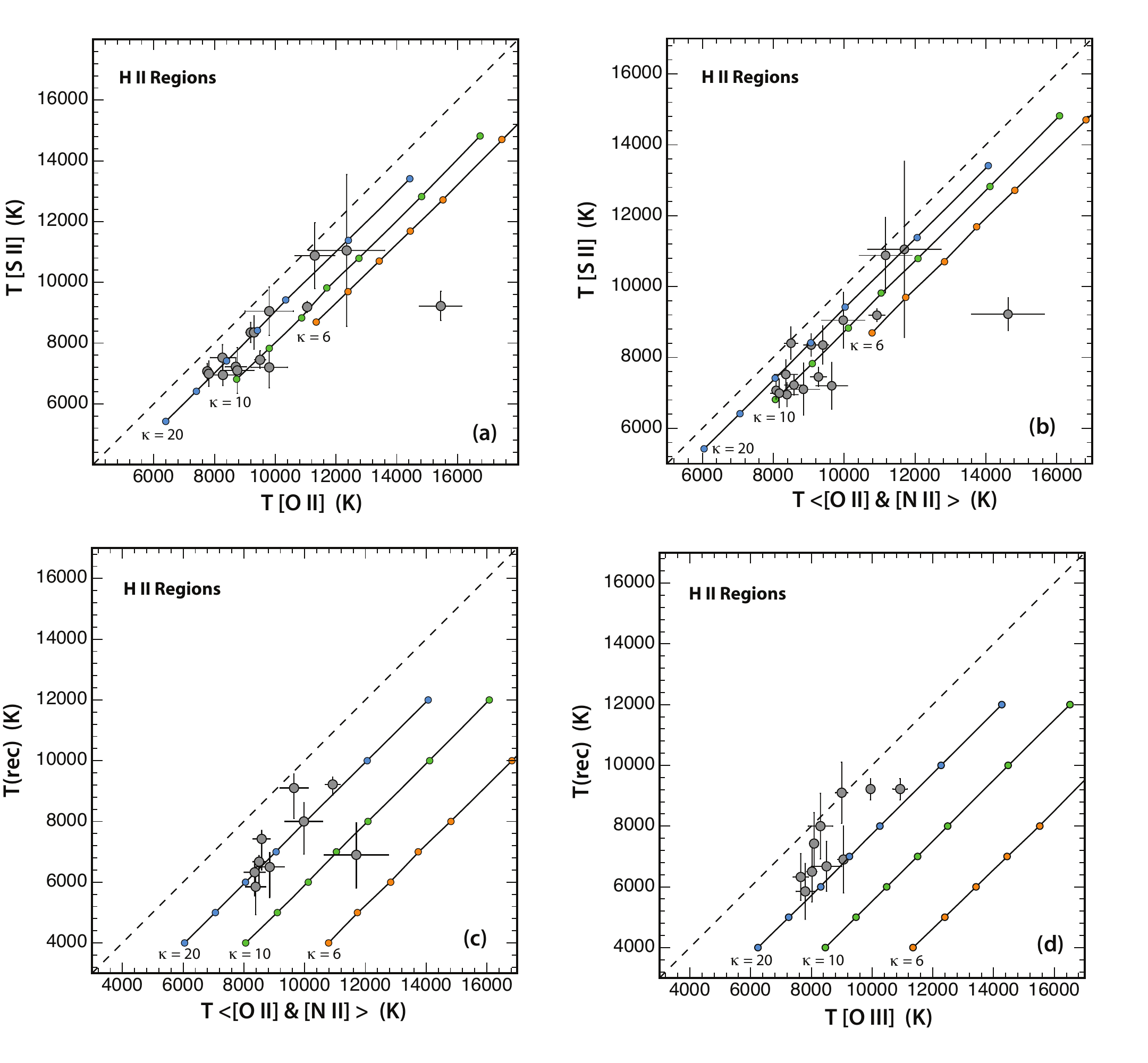}
\caption{Measured temperatures in \HII regions compared with what would be expected if the electrons have a $\kappa$--distribution. Panel (a) compares the inferred excitation temperatures of the  [\ion{O}{2}]  and   [\ion{S}{2}] ions, panel (b) the average of the   [\ion{O}{2}]  and  [\ion{N}{2}] ions with the   [\ion{S}{2}]  ion. Panels (c) and (d) compare the recombination temperatures given by the mean of the Balmer and the Paschen break temperatures with the mean excitation temperature given by the average of the   [\ion{O}{2}]  and  [\ion{N}{2}] line ratios, panel (c), and that of the [\ion{O}{3}] line ratio, panel (d). Most \HII regions appear to have $20 \gtrsim \kappa  \gtrsim10$.}\label{fig_9}
\end{figure*}
We have computed for the observed data the effect that a  $\kappa$--distribution has on the inferred collisional excitation temperature,  $T_{\rm CEL}$, of these ions. This is shown in Figure \ref{fig_9} (a). The observations suggest that the [O II] temperatures are indeed higher than the [S II] temperatures, and that this difference can be accounted for  in most of the \HII regions by $20 \gtrsim \kappa  \gtrsim10$.

%\FloatBarrier % keep images together
As an alterative approach, we show on Figure \ref{fig_9} (b) the average of the inferred collisional excitation temperatures for the  [\ion{O}{2}]  and  [\ion{N}{2}] ions compared with that of the  [\ion{S}{2}]  ion. For the  [\ion{O}{2}]  and  [\ion{N}{2}] ions the sensitivity to $\kappa$ is not as great as for the  [\ion{O}{2}]  lines alone, but the errors in the temperature determination produced by assumptions about the dust extinction curve are reduced. Again, the [S II] temperatures are lower. The inferred range of $\kappa$ is somewhat wider, but most  \HII regions are still consistent with $20 \gtrsim \kappa  \gtrsim10$.

Because the recombination temperature, $T_{\rm B}$, is appreciably lower than the kinetic temperature $T_{\rm U}$, which is itself lower than the excitation temperature for collisionally excited line ratios, $T_{\rm ex}$, a comparison of recombination temperatures and inferred line temperatures is strongly sensitive to the choice of $\kappa$. The recombination temperatures,  $T_{\rm REC}$, should be significantly lower than the collisionally excited line temperatures,  $T_{\rm CEL}$, in the presence of a $\kappa$--distribution.  Again, this is clearly shown in the analysis of the observed \HII region spectra.

In Figures \ref{fig_9} (c) and (d) we show this comparison for the \HII regions, using the excitation temperature given by the average of the   [\ion{O}{2}]  and  [\ion{N}{2}] line ratios and by the [\ion{O}{3}]  line ratio, respectively. How well these determine $\kappa$ depends on which of the ionic species is the dominant ionization stage in the nebula. For cooler exciting stars, [\ion{O}{2}]  and  [\ion{N}{2}] line ratios are the better ones to use, but when the central star(s) is hot enough to ionize helium to \ion{He}{2} in the bulk of the nebula, then the [\ion{O}{3}]  line ratio is the better one to use.
From Figure \ref{fig_9} (c) and (d) it would appear that $\kappa \sim 20$, or even higher.  Because the upper state of [OIII] has the highest excitation temperature, it is the most sensitive of all the ions to the high energy electrons, and therefore lower values of $\kappa$ have greater effects.

\subsection{ Planetary Nebulae}
High quality spectrophotometry also exists for many PNe, although published values of electron temperatures for many ions is rather sparse. However, rather complete data exist for temperatures determined from the Balmer discontinuity and from the [\ion{O}{3}]  line ratio. The direct comparison of these should be a fairly reliable means of estimating $\kappa$, since the \ion{O}{3} ion is most often the dominant ionization stage in PNe.

\begin{figure}[ht]
\includegraphics[width=\hsize]{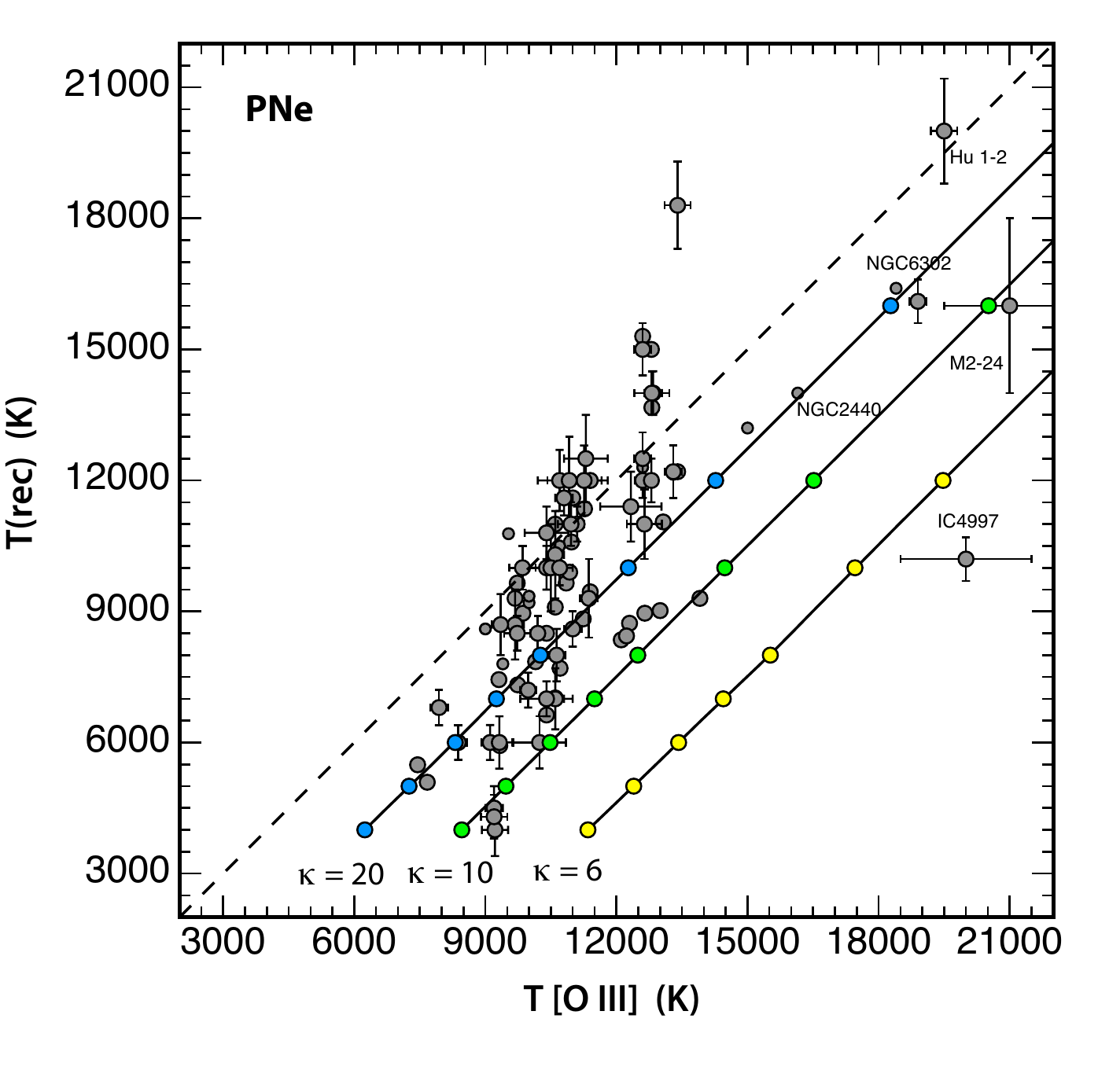}
\caption{As figure 9, panel (d), but for planetary nebulae.}\label{fig_10}
\end{figure}

We have taken data from \citet{Tsamis03,Liu04, Zhang04,Wesson05,Wang07} and \citet{Fang11}. These data are plotted in Figure \ref{fig_10}, along with the expectation for $\kappa = 6, 10$ and 20. The points with error bars are from \citet{Zhang04}.  Again the results are consistent with $\kappa > 10$ in most cases, i.e. a mild departure from an equilibrium electron energy distribution. A few points lie above the line of equality of temperatures shown by the dotted line, corresponding to $\kappa = \infty$. However, many of these are consistent with no departure from a M-B distribution, within the observational errors.  Of course, it is likely that local temperature and density fluctuations in the PNe contribute to the measured results, and it is not obvious what the contributions are from the various effects.

There are two apparent apparent trends in Figure 10: both cooler objects and objects with high [OIII] electron temperatures tend to show greater deviations from equilibrium (smaller inferred $\kappa$). For high electron temperature objects, the energetic stellar winds from the hot central star are likely to lead to significant suprathermal heating of the ionized plasma, leading to lower values of $\kappa$. Indeed, some of the points identified by name (NGC~2440 and NGC~6302) are famous Peimbert Type I  PNe, known to contain the most luminous and massive central stars. Among the rest, IC~4997 is a young PNe which also contains a Wolf-Rayet type central star, subject to rapid mass-loss. 

The coolest objects tend to be more metal rich, and/or of lower excitation class. As a result, the excitation is dominated by the high energy tail of the energy distribution, as illustrated in Figure 5: $\kappa$ effects for [OIII] become more apparent.  These and other characteristics of the $\kappa$--distribution will be the subject of detailed modelling in future papers.

\FloatBarrier % keep images together

The results for both PNe and \HII regions are consistent with mild departures from equilibrium ($\kappa  \gtrsim 10$). It is interesting to note that in the analysis of the observed spectra, extreme values of $\kappa$ are not required. The implied departures from a M-B distribution are quite small, but their effect on the nebular diagnostics is quite gross, with serious implications for the chemical abundance determinations in both classes of object. Large departures from the M-B distribution in either PNe or \HII regions (lower values of $\kappa$) would in any case not be expected, since at the peak of the energy distribution the mean collision time between electrons is short.

\section{Conclusion}

In this paper we have explored the implications for temperature and metallicity measurement in \HII regions and PNe of assuming a non-equilibrium $\kappa$-- electron energy distribution.  We have shown that $\kappa$--distributions provide apparent electron temperatures measured from forbidden line ratios which are systematically higher than the kinetic temperature $T_U$.  Assuming $\kappa$--distributed electron energies appears to resolve the long standing discrepancies in temperatures measured using the collisional excitation lines of different atomic species, and those calculated using the bound-free recombination continuum. Using high quality published spectra, we have shown that  for objects where the spectra of more than one appropriate atomic species is identifiable, it is possible to estimate a value for the effective value of $\kappa$.  $\kappa \sim$ 20 is a good fit to many of the measurements.  This value does not require a substantial redistribution of the population to be achieved, meaning that any of several means of generating high energy electrons should be capable of sustaining the non-equilibrium distribution.

The $\kappa$--distribution offers an important new insight into the physics of gaseous nebulae, both in the Galaxy and elsewhere.  It is implausible that thermal equilibrium applies throughout in \HII regions and PNe, and the fact that the $\kappa$--distribution is able to explain long standing discrepancies is a strong indication that non-equilibrium is a valid assumption.  It should enable more accurate estimates of temperature and metallicity in these regions.  A detailed investigation into the implications for metallicity measurements will be provided in a future paper.

\begin{acknowledgments}
The authors would like to thank the anonymous referee for helpful suggestions on the document structure.  M. Dopita acknowledges the support of the Australian Research Council (ARC) through Discovery  project DP0984657. 
\end{acknowledgments}

\appendix
\section{A.  Evidence for magnetic fields in Milky Way \HII regions}

In order for magnetic energy to be an important source of non-thermal electrons, we require $\beta$, the ratio of the thermal to magnetic energy, to be less than unity. Direct measurements of $\beta$ have been made for \HII regions using Faraday rotation observations \citep{HarveySmith11,Rodriguez12}. These reveal values of $\beta \sim 5$. However, these measurements do not reveal the full strength of the magnetic field if there is turbulence, they indicate only the size of the organised field. They should therefore be taken as imposing a lower bound on the true magnetic field. However, recent evidence on the warm ionized medium \citep{Gaensler11} suggests that magnetic turbulence dominates at small scales. Finally, it is generally believed that the magnetic turbulence in the \HI phase leads to equipartition ($\beta \sim 1$). Let us take this as a working hypothesis and ask what happens across an ionization front in an \HII region.

Let the \HI region be characterized by a density $\rho_0$, temperature $T_0$, pressure $P_0$, magnetic field $\overrightarrow B_0$ and sound speed $c_0 = (\gamma P_0/\rho_0)^{1/2}$.  The Alfven velocity in the gas is $v_A^2 = B_0^2/4\pi\rho$. In the \HI region ahead of the ionization front, if we assume that the gas is turbulently supported, $\beta_0 \sim 1$, \emph{i.e.} $(2/\gamma)(c_0/v_A)^2 \sim 1$, on average. We can resolve the magnetic field into components parallel to the gas flow, $B^{\parallel}_0 \sim B_0 / \sqrt {3}$ and perpendicular to the gas flow (in the plane of the ionization front), $B^{\perp}_0 \sim B_0\sqrt {2/3}$. These components will fluctuate on the scale of the pressure-supported turbulent cells in the \HI region, so they relate to the mean field only in a stochastic fashion.

Now consider what happens on the other side of the ionization front, where the sound speed is $c_1$, the outflow Mach number is ${\mathcal M}$ and the hydrodynamic variables are $\rho_1,~P_1,~ T_1, ~B^{\parallel}_1$ and $B^{\perp}_1$. The ratio of the densities is obtained by equating the pre-ionization pressure to the post-ionization pressure, accounting for the recoil momentum of the ionized plasma;
\begin{equation}\label{eqA1}
{\frac{\rho_o c_o^2}{ \gamma}}  =\frac{\rho_1 c_1^2 }{ \gamma} + {\mathcal M }^2\rho_1c_1^2\ .
\end{equation}
The component of the magnetic field perpendicular to the flow direction is stretched (reduced) in the post-ionization zone by the ratio $\rho_1/\rho_0$, and the local gas pressure is decreased in the ratio $1/(1+ \gamma {\mathcal M }^2)$. Therefore:
\begin{equation}\label{eqA2}
\beta^{\perp}_1 \sim {\frac{3}{ 2}}\left({\frac{1}{1+ \gamma {\mathcal M }^2}} \right ) \left({\frac{ \rho_0}{\rho_1}}\right)^2\ .
\end{equation}
Thus, all magnetic pressure support is effectively lost in this plane. However, parallel to the flow, the post-shock magnetic field is unchanged by the ionization, and therefore
\begin{equation}\label{eqA3}
\beta^{\parallel}_1 \sim {3}\left({\frac{1}{ 1+ \gamma {\mathcal M }^2}} \right ) < 1\ .
\end{equation}
Thus, the magnetic pressure dominates the thermal pressure in this direction. The turbulent origin of the magnetic field in the molecular cloud ensures that $\beta^{\parallel}_1$ is rapidly fluctuating both in magnitude and direction, which will naturally assist magnetic reconnection and Inertial Alfv\'en Wave formation. 

\begin{figure*}[h] % Figure 11 full width
\includegraphics[width=\hsize]{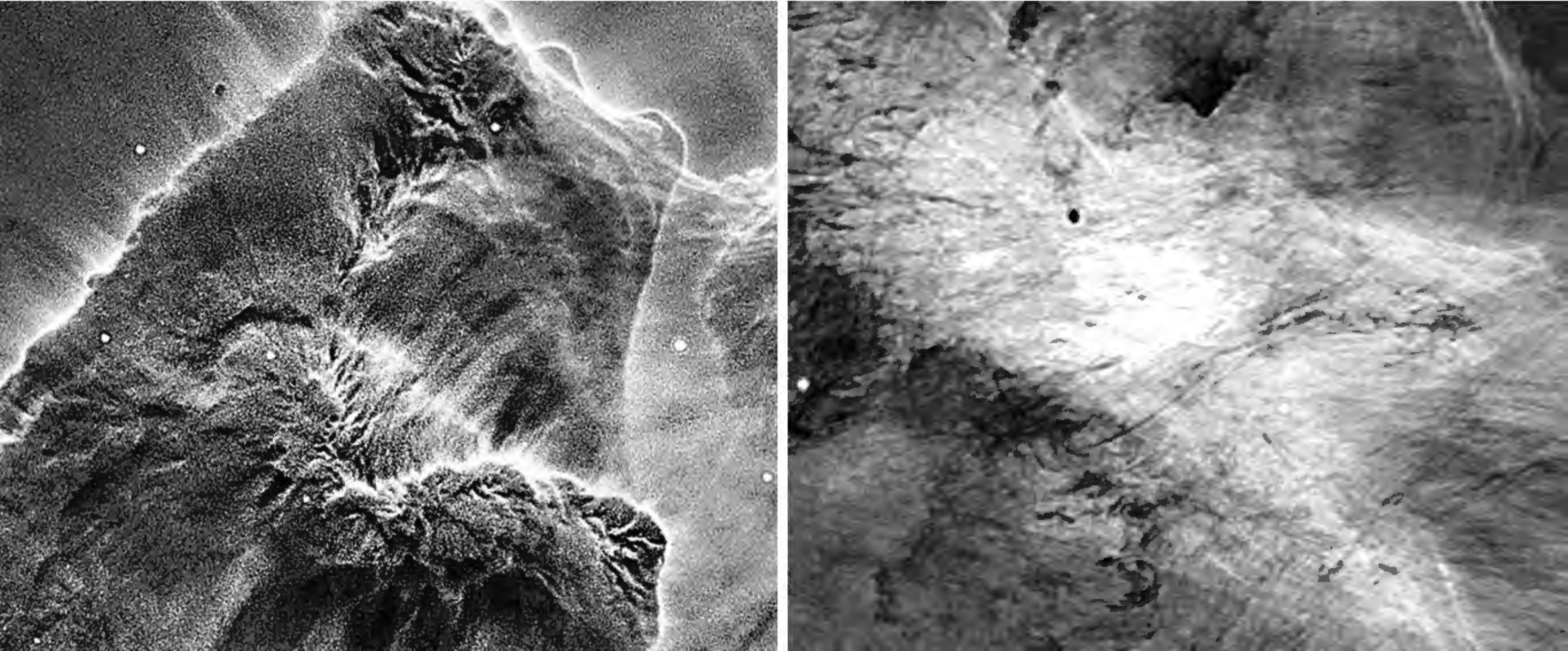}
\caption{Portions of HST images of  the $\eta$ Carina nebula (left) and M17 (right) taken with the ACS. These images have been processed by an unsharp mask to bring up the microstructure. Both images display magnetically-dominated microstructures aligned with the outflow direction. The scales of these microfilaments is of order of the size of the solar system.}\label{fig_11}
\end{figure*}

This magnetic field, highly aligned to the flow direction and showing strong fluctuations on the small scale provides a natural explanation for the filamentary structure of the ionized plasma seen very clearly in the $\eta$ Carina and M17 \HII regions observed by the Advanced Camera for Surveys (ACS) on the Hubble Space Telescope (HST) (see Figure \ref{fig_11}, below). 

In the case of  $\eta$ Carina, the ionized filaments are seen as bright ``hairs'' originating at the ionization front. These bright regions correspond to the regions of low magnetic field, given that the gas flow from the ionization front has to be force free in the plane of the ionization front. In the case of M17, we are seeing un-ionized regions of low magnetic field and high dust optical depth ``combed out'' by the expanding ionized plasma. These filaments are only a few AU in diameter, and have an aspect ratio of up to $50:1$. The implied atomic densities are of order $10^5$ atoms cm$^{-3}$.\\

\emph{Note added in proof} Our attention has been drawn to simulations of the magnetohydrodynamics of \HII regions by \citet{Henney09} and \citet{Arthur11}, in which they show that, in their models, $\beta$, the ratio of thermal to magnetic energy, is usually greater than unity, and that the observed filamentary structures (Figure 11) arise naturally in the simulations.  The fact that these features appear to be magnetohydrodynamic in origin is evidence for the widespread presence of magnetic fields in \HII regions.

\newpage
\ \\ \\
 \section{B.  The $\kappa$--distribution and the temperature fluctuation (``$t^2$'') method}

As early as 1967, Peimbert proposed that abundance discrepancies might be explained by spatial temperature variations within an \HII region \citep{Peimbert67}.  The approach is characterized by a parameter, $t^2$, the mean square fractional temperature fluctuation (assumed to have a Gaussian distribution in space). While it is certain that there are spatial variations in temperature, those caused naturally by the radial temperature variation in the \HII region are usually insufficient to explain observations, and the cause of the larger values of  $t^2$ implied by the observations remains unknown. However, the $t^2$ approach has been widely used, with mixed success (e.g. \citet{Pena12,G-R07,Esteban02,Kingdon95}).  Detailed discussions of the technique are given in \citet{Stasinska04} and \citet{Kingdon95}.

The $t^2$ parameter is estimated by solving equations involving the collisionally excited line temperature(s) and the recombination line temperature \citep{Kingdon95}.  But as \citet{Kingdon95} also argued, ``while temperature fluctuations may result in non-negligible abundance corrections in some objects, they are insufficient to resolve the abundance discrepancy''.

There is, however, a connection between the $t^2$ method for resolving abundance discrepancies and the $\kappa$--distribution.  \citet{Livadiotis11a} noted that ``$\kappa$--distributions can be derived from the superposition of Maxwellian distributions, by considering a non-fixed temperature that is characterised by a certain (continuous or discrete) density distribution of temperatures''.  Thus:
\begin{equation}\label{eq19}
f_\kappa(E,T_U,\kappa) = \sum_{i=0}^{\infty}\Phi(f_{Max}(E,T_i),\kappa)\ ,
\end{equation}
where $\Phi$ can, in theory, be determined from the expressions in equations 6 and 7.  The necessity to include Maxwellian distributions in the series with extreme temperatures to provide the suprathermal tail at high energies, is, of course, non-physical, and the series needs to be truncated. Nonetheless, expressed this way, the $\kappa$--distribution can be seen as a generalization of the $t^2$ method.

\newpage

\begin{center}
\begin{longtable*}{rrrrrrrrr}
%Here is the caption, the stuff in [] is the table of contents entry,
%the stuff in {} is the title that will appear on the first page of the
%table.
\caption[Collisional Excitation Rate ratios (see Fig. 6)]{\sc Collisional Excitation Rate ratios (see Fig. 5)} \label{table3} \\

%This is the header for the first page of the table...
\hline \hline \\
\multicolumn{1}{c}{$T_{ex}/T_U$} & 
\multicolumn{1}{c}{$\kappa=2$} & 
\multicolumn{1}{c}{$\kappa=3$} & 
\multicolumn{1}{c}{$\kappa=4$} & 
\multicolumn{1}{c}{$\kappa=6$} & 
\multicolumn{1}{c}{$\kappa=10$} & 
\multicolumn{1}{c}{$\kappa=20$} & 
\multicolumn{1}{c}{$\kappa=50$} & 
\multicolumn{1}{c}{$\kappa=100$} \\[0.5ex] \hline
  \\
\endfirsthead

%This is the header for the remaining page(s) of the table...
\multicolumn{9}{c}{{\tablename} \thetable{} -- Continued} \\[0.5ex]
 \hline \hline \\
\multicolumn{1}{c}{$T_{ex}/T_U$} & 
\multicolumn{1}{c}{$\kappa=2$} & 
\multicolumn{1}{c}{$\kappa=3$} & 
\multicolumn{1}{c}{$\kappa=4$} & 
\multicolumn{1}{c}{$\kappa=6$} & 
\multicolumn{1}{c}{$\kappa=10$} & 
\multicolumn{1}{c}{$\kappa=20$} & 
\multicolumn{1}{c}{$\kappa=50$} & 
\multicolumn{1}{c}{$\kappa=100$} \\ \hline
  \\ 
\endhead

%This is the footer for all pages except the last page of the table...
  \multicolumn{9}{l}{{\ldots next}} \\
\endfoot

%This is the footer for the last page of the table...
  \\ \hline \hline
\endlastfoot
%Now the data...

0.0 & 1.596E+00 & 1.228E+00 & 1.142E+00 & 1.081E+00 & 1.043E+00 & 1.020E+00 & 1.008E+00 & 1.004E+00 \\ 
0.1 & 1.225E+00 & 1.119E+00 & 1.079E+00 & 1.047E+00 & 1.026E+00 & 1.012E+00 & 1.005E+00 & 1.002E+00 \\ 
0.2 & 9.944E-01 & 1.031E+00 & 1.025E+00 & 1.017E+00 & 1.010E+00 & 1.005E+00 & 1.002E+00 & 1.001E+00 \\ 
0.3 & 8.414E-01 & 9.596E-01 & 9.795E-01 & 9.905E-01 & 9.956E-01 & 9.982E-01 & 9.994E-01 & 9.997E-01 \\ 
0.4 & 7.348E-01 & 9.017E-01 & 9.408E-01 & 9.672E-01 & 9.828E-01 & 9.921E-01 & 9.970E-01 & 9.985E-01 \\ 
0.5 & 6.577E-01 & 8.544E-01 & 9.079E-01 & 9.469E-01 & 9.713E-01 & 9.866E-01 & 9.949E-01 & 9.975E-01 \\ 
0.6 & 6.008E-01 & 8.157E-01 & 8.800E-01 & 9.293E-01 & 9.612E-01 & 9.817E-01 & 9.929E-01 & 9.965E-01 \\ 
0.7 & 5.579E-01 & 7.841E-01 & 8.566E-01 & 9.141E-01 & 9.523E-01 & 9.774E-01 & 9.912E-01 & 9.957E-01 \\ 
0.8 & 5.254E-01 & 7.584E-01 & 8.370E-01 & 9.011E-01 & 9.446E-01 & 9.736E-01 & 9.897E-01 & 9.949E-01 \\ 
0.9 & 5.006E-01 & 7.377E-01 & 8.209E-01 & 8.902E-01 & 9.380E-01 & 9.703E-01 & 9.884E-01 & 9.943E-01 \\ 
1.0 & 4.820E-01 & 7.213E-01 & 8.080E-01 & 8.813E-01 & 9.326E-01 & 9.676E-01 & 9.873E-01 & 9.937E-01 \\ 
1.1 & 4.682E-01 & 7.086E-01 & 7.978E-01 & 8.742E-01 & 9.282E-01 & 9.654E-01 & 9.864E-01 & 9.933E-01 \\ 
1.2 & 4.583E-01 & 6.993E-01 & 7.902E-01 & 8.688E-01 & 9.248E-01 & 9.636E-01 & 9.857E-01 & 9.929E-01 \\ 
1.3 & 4.518E-01 & 6.930E-01 & 7.849E-01 & 8.650E-01 & 9.225E-01 & 9.624E-01 & 9.852E-01 & 9.927E-01 \\ 
1.4 & 4.481E-01 & 6.894E-01 & 7.818E-01 & 8.628E-01 & 9.211E-01 & 9.617E-01 & 9.849E-01 & 9.925E-01 \\ 
1.5 & 4.470E-01 & 6.882E-01 & 7.808E-01 & 8.620E-01 & 9.206E-01 & 9.614E-01 & 9.848E-01 & 9.925E-01 \\ 
1.6 & 4.481E-01 & 6.893E-01 & 7.818E-01 & 8.628E-01 & 9.211E-01 & 9.617E-01 & 9.849E-01 & 9.925E-01 \\ 
1.7 & 4.512E-01 & 6.926E-01 & 7.846E-01 & 8.649E-01 & 9.224E-01 & 9.624E-01 & 9.852E-01 & 9.927E-01 \\ 
1.8 & 4.562E-01 & 6.979E-01 & 7.893E-01 & 8.683E-01 & 9.247E-01 & 9.636E-01 & 9.857E-01 & 9.929E-01 \\ 
1.9 & 4.631E-01 & 7.053E-01 & 7.956E-01 & 8.731E-01 & 9.278E-01 & 9.652E-01 & 9.864E-01 & 9.933E-01 \\ 
2.0 & 4.716E-01 & 7.145E-01 & 8.037E-01 & 8.792E-01 & 9.318E-01 & 9.674E-01 & 9.873E-01 & 9.937E-01 \\ 
2.1 & 4.819E-01 & 7.257E-01 & 8.135E-01 & 8.866E-01 & 9.367E-01 & 9.700E-01 & 9.884E-01 & 9.942E-01 \\ 
2.2 & 4.939E-01 & 7.387E-01 & 8.249E-01 & 8.953E-01 & 9.424E-01 & 9.730E-01 & 9.896E-01 & 9.949E-01 \\ 
2.3 & 5.075E-01 & 7.536E-01 & 8.381E-01 & 9.054E-01 & 9.490E-01 & 9.765E-01 & 9.911E-01 & 9.956E-01 \\ 
2.4 & 5.229E-01 & 7.704E-01 & 8.529E-01 & 9.167E-01 & 9.565E-01 & 9.805E-01 & 9.927E-01 & 9.965E-01 \\ 
2.5 & 5.400E-01 & 7.892E-01 & 8.694E-01 & 9.293E-01 & 9.648E-01 & 9.850E-01 & 9.946E-01 & 9.974E-01 \\ 
2.6 & 5.589E-01 & 8.099E-01 & 8.877E-01 & 9.432E-01 & 9.740E-01 & 9.899E-01 & 9.966E-01 & 9.984E-01 \\ 
2.7 & 5.797E-01 & 8.327E-01 & 9.077E-01 & 9.585E-01 & 9.841E-01 & 9.953E-01 & 9.989E-01 & 9.996E-01 \\ 
2.8 & 6.024E-01 & 8.575E-01 & 9.296E-01 & 9.752E-01 & 9.951E-01 & 1.001E+00 & 1.001E+00 & 1.001E+00 \\ 
2.9 & 6.272E-01 & 8.845E-01 & 9.533E-01 & 9.932E-01 & 1.007E+00 & 1.008E+00 & 1.004E+00 & 1.002E+00 \\ 
3.0 & 6.541E-01 & 9.138E-01 & 9.790E-01 & 1.013E+00 & 1.020E+00 & 1.014E+00 & 1.007E+00 & 1.004E+00 \\ 
3.1 & 6.833E-01 & 9.455E-01 & 1.007E+00 & 1.034E+00 & 1.034E+00 & 1.022E+00 & 1.010E+00 & 1.005E+00 \\ 
3.2 & 7.149E-01 & 9.796E-01 & 1.037E+00 & 1.056E+00 & 1.048E+00 & 1.030E+00 & 1.013E+00 & 1.007E+00 \\ 
3.3 & 7.491E-01 & 1.016E+00 & 1.069E+00 & 1.080E+00 & 1.064E+00 & 1.038E+00 & 1.017E+00 & 1.009E+00 \\ 
3.4 & 7.859E-01 & 1.056E+00 & 1.103E+00 & 1.106E+00 & 1.081E+00 & 1.047E+00 & 1.020E+00 & 1.010E+00 \\ 
3.5 & 8.257E-01 & 1.098E+00 & 1.140E+00 & 1.134E+00 & 1.099E+00 & 1.056E+00 & 1.024E+00 & 1.012E+00 \\ 
3.6 & 8.686E-01 & 1.144E+00 & 1.179E+00 & 1.163E+00 & 1.118E+00 & 1.066E+00 & 1.028E+00 & 1.014E+00 \\ 
3.7 & 9.147E-01 & 1.193E+00 & 1.221E+00 & 1.194E+00 & 1.137E+00 & 1.076E+00 & 1.032E+00 & 1.016E+00 \\ 
3.8 & 9.645E-01 & 1.245E+00 & 1.266E+00 & 1.227E+00 & 1.159E+00 & 1.087E+00 & 1.037E+00 & 1.019E+00 \\ 
3.9 & 1.018E+00 & 1.301E+00 & 1.313E+00 & 1.262E+00 & 1.181E+00 & 1.099E+00 & 1.041E+00 & 1.021E+00 \\ 
4.0 & 1.076E+00 & 1.361E+00 & 1.364E+00 & 1.299E+00 & 1.204E+00 & 1.111E+00 & 1.046E+00 & 1.023E+00 \\ 
4.1 & 1.138E+00 & 1.425E+00 & 1.418E+00 & 1.339E+00 & 1.229E+00 & 1.123E+00 & 1.051E+00 & 1.026E+00 \\ 
4.2 & 1.204E+00 & 1.493E+00 & 1.476E+00 & 1.380E+00 & 1.255E+00 & 1.137E+00 & 1.057E+00 & 1.029E+00 \\ 
4.3 & 1.276E+00 & 1.566E+00 & 1.537E+00 & 1.424E+00 & 1.282E+00 & 1.150E+00 & 1.062E+00 & 1.031E+00 \\ 
4.4 & 1.353E+00 & 1.644E+00 & 1.603E+00 & 1.471E+00 & 1.311E+00 & 1.165E+00 & 1.068E+00 & 1.034E+00 \\ 
4.5 & 1.437E+00 & 1.728E+00 & 1.672E+00 & 1.520E+00 & 1.341E+00 & 1.180E+00 & 1.074E+00 & 1.037E+00 \\ 
4.6 & 1.526E+00 & 1.817E+00 & 1.746E+00 & 1.572E+00 & 1.373E+00 & 1.196E+00 & 1.080E+00 & 1.040E+00 \\ 
4.7 & 1.622E+00 & 1.913E+00 & 1.825E+00 & 1.627E+00 & 1.406E+00 & 1.212E+00 & 1.087E+00 & 1.044E+00 \\ 
4.8 & 1.726E+00 & 2.015E+00 & 1.909E+00 & 1.685E+00 & 1.441E+00 & 1.229E+00 & 1.093E+00 & 1.047E+00 \\ 
4.9 & 1.837E+00 & 2.124E+00 & 1.998E+00 & 1.747E+00 & 1.478E+00 & 1.247E+00 & 1.100E+00 & 1.050E+00 \\ 
5.0 & 1.957E+00 & 2.241E+00 & 2.092E+00 & 1.812E+00 & 1.516E+00 & 1.265E+00 & 1.107E+00 & 1.054E+00 \\ 
5.1 & 2.087E+00 & 2.365E+00 & 2.193E+00 & 1.881E+00 & 1.557E+00 & 1.285E+00 & 1.115E+00 & 1.057E+00 \\ 
5.2 & 2.226E+00 & 2.499E+00 & 2.300E+00 & 1.953E+00 & 1.599E+00 & 1.305E+00 & 1.122E+00 & 1.061E+00 \\ 
5.3 & 2.376E+00 & 2.642E+00 & 2.414E+00 & 2.030E+00 & 1.643E+00 & 1.325E+00 & 1.130E+00 & 1.065E+00 \\ 
5.4 & 2.537E+00 & 2.794E+00 & 2.535E+00 & 2.110E+00 & 1.689E+00 & 1.347E+00 & 1.138E+00 & 1.069E+00 \\ 
5.5 & 2.712E+00 & 2.958E+00 & 2.665E+00 & 2.196E+00 & 1.738E+00 & 1.369E+00 & 1.146E+00 & 1.073E+00 \\ 
5.6 & 2.899E+00 & 3.133E+00 & 2.802E+00 & 2.286E+00 & 1.789E+00 & 1.393E+00 & 1.155E+00 & 1.077E+00 \\ 
5.7 & 3.102E+00 & 3.320E+00 & 2.948E+00 & 2.382E+00 & 1.842E+00 & 1.417E+00 & 1.164E+00 & 1.081E+00 \\ 
5.8 & 3.320E+00 & 3.520E+00 & 3.104E+00 & 2.482E+00 & 1.898E+00 & 1.442E+00 & 1.173E+00 & 1.086E+00 \\ 
5.9 & 3.555E+00 & 3.735E+00 & 3.270E+00 & 2.589E+00 & 1.956E+00 & 1.468E+00 & 1.183E+00 & 1.090E+00 \\ 
6.0 & 3.809E+00 & 3.965E+00 & 3.447E+00 & 2.702E+00 & 2.017E+00 & 1.495E+00 & 1.192E+00 & 1.095E+00 \\ 
6.1 & 4.083E+00 & 4.211E+00 & 3.636E+00 & 2.821E+00 & 2.081E+00 & 1.523E+00 & 1.202E+00 & 1.100E+00 \\ 
6.2 & 4.379E+00 & 4.475E+00 & 3.836E+00 & 2.947E+00 & 2.148E+00 & 1.552E+00 & 1.213E+00 & 1.105E+00 \\ 
6.3 & 4.698E+00 & 4.758E+00 & 4.050E+00 & 3.080E+00 & 2.219E+00 & 1.582E+00 & 1.223E+00 & 1.110E+00 \\ 
6.4 & 5.043E+00 & 5.061E+00 & 4.279E+00 & 3.221E+00 & 2.292E+00 & 1.613E+00 & 1.234E+00 & 1.115E+00 \\ 
6.5 & 5.415E+00 & 5.386E+00 & 4.522E+00 & 3.369E+00 & 2.369E+00 & 1.645E+00 & 1.245E+00 & 1.120E+00 \\ 
6.6 & 5.818E+00 & 5.735E+00 & 4.781E+00 & 3.527E+00 & 2.450E+00 & 1.679E+00 & 1.257E+00 & 1.126E+00 \\ 
6.7 & 6.252E+00 & 6.109E+00 & 5.058E+00 & 3.694E+00 & 2.535E+00 & 1.713E+00 & 1.268E+00 & 1.131E+00 \\ 
6.8 & 6.722E+00 & 6.510E+00 & 5.354E+00 & 3.870E+00 & 2.624E+00 & 1.749E+00 & 1.281E+00 & 1.137E+00 \\ 
6.9 & 7.229E+00 & 6.941E+00 & 5.669E+00 & 4.057E+00 & 2.717E+00 & 1.787E+00 & 1.293E+00 & 1.143E+00 \\ 
7.0 & 7.778E+00 & 7.403E+00 & 6.005E+00 & 4.255E+00 & 2.814E+00 & 1.825E+00 & 1.306E+00 & 1.148E+00 \\ 
7.1 & 8.371E+00 & 7.900E+00 & 6.365E+00 & 4.464E+00 & 2.917E+00 & 1.865E+00 & 1.319E+00 & 1.154E+00 \\ 
7.2 & 9.013E+00 & 8.433E+00 & 6.748E+00 & 4.686E+00 & 3.024E+00 & 1.907E+00 & 1.332E+00 & 1.161E+00 \\ 
7.3 & 9.707E+00 & 9.006E+00 & 7.158E+00 & 4.921E+00 & 3.136E+00 & 1.950E+00 & 1.346E+00 & 1.167E+00 \\ 
7.4 & 1.046E+01 & 9.621E+00 & 7.596E+00 & 5.170E+00 & 3.254E+00 & 1.995E+00 & 1.360E+00 & 1.173E+00 \\ 
7.5 & 1.127E+01 & 1.028E+01 & 8.064E+00 & 5.434E+00 & 3.378E+00 & 2.041E+00 & 1.375E+00 & 1.180E+00 \\ 
7.6 & 1.215E+01 & 1.099E+01 & 8.565E+00 & 5.714E+00 & 3.508E+00 & 2.089E+00 & 1.390E+00 & 1.187E+00 \\ 
7.7 & 1.310E+01 & 1.176E+01 & 9.100E+00 & 6.010E+00 & 3.644E+00 & 2.139E+00 & 1.405E+00 & 1.194E+00 \\ 
7.8 & 1.413E+01 & 1.258E+01 & 9.672E+00 & 6.325E+00 & 3.787E+00 & 2.190E+00 & 1.421E+00 & 1.201E+00 \\ 
7.9 & 1.525E+01 & 1.346E+01 & 1.028E+01 & 6.659E+00 & 3.937E+00 & 2.244E+00 & 1.437E+00 & 1.208E+00 \\ 
8.0 & 1.646E+01 & 1.441E+01 & 1.094E+01 & 7.013E+00 & 4.094E+00 & 2.299E+00 & 1.453E+00 & 1.215E+00 \\ 
8.1 & 1.777E+01 & 1.544E+01 & 1.164E+01 & 7.389E+00 & 4.259E+00 & 2.356E+00 & 1.470E+00 & 1.223E+00 \\ 
8.2 & 1.919E+01 & 1.654E+01 & 1.239E+01 & 7.787E+00 & 4.433E+00 & 2.416E+00 & 1.488E+00 & 1.230E+00 \\ 
8.3 & 2.073E+01 & 1.773E+01 & 1.319E+01 & 8.211E+00 & 4.615E+00 & 2.478E+00 & 1.505E+00 & 1.238E+00 \\ 
8.4 & 2.240E+01 & 1.900E+01 & 1.405E+01 & 8.660E+00 & 4.807E+00 & 2.542E+00 & 1.524E+00 & 1.246E+00 \\ 
8.5 & 2.421E+01 & 2.038E+01 & 1.497E+01 & 9.138E+00 & 5.008E+00 & 2.608E+00 & 1.542E+00 & 1.254E+00 \\ 
8.6 & 2.617E+01 & 2.186E+01 & 1.596E+01 & 9.645E+00 & 5.219E+00 & 2.677E+00 & 1.562E+00 & 1.263E+00 \\ 
8.7 & 2.829E+01 & 2.345E+01 & 1.702E+01 & 1.018E+01 & 5.441E+00 & 2.748E+00 & 1.581E+00 & 1.271E+00 \\ 
8.8 & 3.060E+01 & 2.517E+01 & 1.815E+01 & 1.076E+01 & 5.675E+00 & 2.822E+00 & 1.601E+00 & 1.280E+00 \\ 
8.9 & 3.310E+01 & 2.702E+01 & 1.936E+01 & 1.137E+01 & 5.921E+00 & 2.899E+00 & 1.622E+00 & 1.288E+00 \\ 
9.0 & 3.582E+01 & 2.902E+01 & 2.066E+01 & 1.201E+01 & 6.179E+00 & 2.979E+00 & 1.643E+00 & 1.297E+00 \\ 
9.1 & 3.877E+01 & 3.117E+01 & 2.206E+01 & 1.270E+01 & 6.450E+00 & 3.062E+00 & 1.665E+00 & 1.307E+00 \\ 
9.2 & 4.196E+01 & 3.350E+01 & 2.356E+01 & 1.343E+01 & 6.736E+00 & 3.148E+00 & 1.687E+00 & 1.316E+00 \\ 
9.3 & 4.544E+01 & 3.600E+01 & 2.516E+01 & 1.421E+01 & 7.037E+00 & 3.237E+00 & 1.710E+00 & 1.325E+00 \\ 
9.4 & 4.920E+01 & 3.870E+01 & 2.689E+01 & 1.504E+01 & 7.353E+00 & 3.329E+00 & 1.733E+00 & 1.335E+00 \\ 
9.5 & 5.330E+01 & 4.161E+01 & 2.874E+01 & 1.592E+01 & 7.686E+00 & 3.426E+00 & 1.757E+00 & 1.345E+00 \\ 
9.6 & 5.774E+01 & 4.476E+01 & 3.072E+01 & 1.686E+01 & 8.037E+00 & 3.525E+00 & 1.782E+00 & 1.355E+00 \\ 
9.7 & 6.257E+01 & 4.815E+01 & 3.285E+01 & 1.786E+01 & 8.406E+00 & 3.629E+00 & 1.807E+00 & 1.365E+00 \\ 
9.8 & 6.781E+01 & 5.182E+01 & 3.514E+01 & 1.893E+01 & 8.794E+00 & 3.736E+00 & 1.833E+00 & 1.376E+00 \\ 
9.9 & 7.351E+01 & 5.577E+01 & 3.760E+01 & 2.006E+01 & 9.204E+00 & 3.848E+00 & 1.859E+00 & 1.386E+00 \\ 
10.0 & 7.970E+01 & 6.004E+01 & 4.024E+01 & 2.127E+01 & 9.635E+00 & 3.964E+00 & 1.886E+00 & 1.397E+00 \\ 
10.1 & 8.643E+01 & 6.466E+01 & 4.308E+01 & 2.256E+01 & 1.009E+01 & 4.085E+00 & 1.914E+00 & 1.408E+00 \\ 
10.2 & 9.374E+01 & 6.964E+01 & 4.613E+01 & 2.393E+01 & 1.057E+01 & 4.210E+00 & 1.942E+00 & 1.420E+00 \\ 
10.3 & 1.017E+02 & 7.503E+01 & 4.940E+01 & 2.539E+01 & 1.107E+01 & 4.340E+00 & 1.972E+00 & 1.431E+00 \\ 
10.4 & 1.103E+02 & 8.084E+01 & 5.293E+01 & 2.695E+01 & 1.161E+01 & 4.475E+00 & 2.001E+00 & 1.443E+00 \\ 
10.5 & 1.197E+02 & 8.713E+01 & 5.671E+01 & 2.861E+01 & 1.217E+01 & 4.616E+00 & 2.032E+00 & 1.455E+00 \\ 
10.6 & 1.300E+02 & 9.393E+01 & 6.079E+01 & 3.038E+01 & 1.276E+01 & 4.762E+00 & 2.063E+00 & 1.467E+00 \\ 
10.7 & 1.411E+02 & 1.013E+02 & 6.517E+01 & 3.228E+01 & 1.338E+01 & 4.913E+00 & 2.096E+00 & 1.479E+00 \\ 
10.8 & 1.532E+02 & 1.092E+02 & 6.988E+01 & 3.429E+01 & 1.404E+01 & 5.071E+00 & 2.129E+00 & 1.492E+00 \\ 
10.9 & 1.663E+02 & 1.178E+02 & 7.495E+01 & 3.645E+01 & 1.474E+01 & 5.235E+00 & 2.162E+00 & 1.504E+00 \\ 
11.0 & 1.806E+02 & 1.271E+02 & 8.040E+01 & 3.875E+01 & 1.547E+01 & 5.406E+00 & 2.197E+00 & 1.517E+00 \\ 
11.1 & 1.962E+02 & 1.371E+02 & 8.627E+01 & 4.120E+01 & 1.625E+01 & 5.584E+00 & 2.232E+00 & 1.531E+00 \\ 
11.2 & 2.131E+02 & 1.480E+02 & 9.259E+01 & 4.382E+01 & 1.706E+01 & 5.768E+00 & 2.269E+00 & 1.544E+00 \\ 
11.3 & 2.316E+02 & 1.598E+02 & 9.940E+01 & 4.662E+01 & 1.793E+01 & 5.961E+00 & 2.306E+00 & 1.558E+00 \\ 
11.4 & 2.516E+02 & 1.725E+02 & 1.067E+02 & 4.961E+01 & 1.884E+01 & 6.161E+00 & 2.344E+00 & 1.572E+00 \\ 
11.5 & 2.735E+02 & 1.863E+02 & 1.146E+02 & 5.280E+01 & 1.980E+01 & 6.369E+00 & 2.384E+00 & 1.586E+00 \\ 
11.6 & 2.973E+02 & 2.012E+02 & 1.231E+02 & 5.622E+01 & 2.082E+01 & 6.585E+00 & 2.424E+00 & 1.601E+00 \\ 
11.7 & 3.232E+02 & 2.173E+02 & 1.323E+02 & 5.986E+01 & 2.190E+01 & 6.811E+00 & 2.465E+00 & 1.616E+00 \\ 
11.8 & 3.514E+02 & 2.348E+02 & 1.421E+02 & 6.376E+01 & 2.303E+01 & 7.045E+00 & 2.507E+00 & 1.631E+00 \\ 
11.9 & 3.821E+02 & 2.538E+02 & 1.528E+02 & 6.793E+01 & 2.423E+01 & 7.290E+00 & 2.551E+00 & 1.646E+00 \\ 
12.0 & 4.156E+02 & 2.743E+02 & 1.642E+02 & 7.238E+01 & 2.551E+01 & 7.544E+00 & 2.595E+00 & 1.662E+00 \\ 
12.1 & 4.520E+02 & 2.965E+02 & 1.766E+02 & 7.715E+01 & 2.685E+01 & 7.809E+00 & 2.641E+00 & 1.678E+00 \\ 
12.2 & 4.917E+02 & 3.205E+02 & 1.899E+02 & 8.224E+01 & 2.827E+01 & 8.086E+00 & 2.688E+00 & 1.694E+00 \\ 
12.3 & 5.349E+02 & 3.466E+02 & 2.042E+02 & 8.769E+01 & 2.977E+01 & 8.373E+00 & 2.736E+00 & 1.711E+00 \\ 
12.4 & 5.821E+02 & 3.748E+02 & 2.197E+02 & 9.352E+01 & 3.136E+01 & 8.673E+00 & 2.785E+00 & 1.727E+00 \\ 
12.5 & 6.334E+02 & 4.054E+02 & 2.364E+02 & 9.977E+01 & 3.305E+01 & 8.985E+00 & 2.835E+00 & 1.744E+00 \\ 
12.6 & 6.894E+02 & 4.386E+02 & 2.544E+02 & 1.064E+02 & 3.483E+01 & 9.311E+00 & 2.887E+00 & 1.762E+00 \\ 
12.7 & 7.504E+02 & 4.746E+02 & 2.739E+02 & 1.136E+02 & 3.671E+01 & 9.650E+00 & 2.940E+00 & 1.780E+00 \\ 
12.8 & 8.169E+02 & 5.136E+02 & 2.948E+02 & 1.213E+02 & 3.871E+01 & 1.000E+01 & 2.995E+00 & 1.798E+00 \\ 
12.9 & 8.894E+02 & 5.558E+02 & 3.175E+02 & 1.294E+02 & 4.082E+01 & 1.037E+01 & 3.051E+00 & 1.816E+00 \\ 
13.0 & 9.684E+02 & 6.017E+02 & 3.419E+02 & 1.382E+02 & 4.306E+01 & 1.076E+01 & 3.108E+00 & 1.835E+00 \\ 
13.1 & 1.055E+03 & 6.514E+02 & 3.682E+02 & 1.476E+02 & 4.543E+01 & 1.116E+01 & 3.167E+00 & 1.854E+00 \\ 
13.2 & 1.149E+03 & 7.053E+02 & 3.967E+02 & 1.577E+02 & 4.794E+01 & 1.158E+01 & 3.227E+00 & 1.873E+00 \\ 
13.3 & 1.251E+03 & 7.638E+02 & 4.274E+02 & 1.685E+02 & 5.060E+01 & 1.201E+01 & 3.289E+00 & 1.893E+00 \\ 
13.4 & 1.363E+03 & 8.272E+02 & 4.606E+02 & 1.801E+02 & 5.342E+01 & 1.247E+01 & 3.353E+00 & 1.913E+00 \\ 
13.5 & 1.485E+03 & 8.960E+02 & 4.964E+02 & 1.925E+02 & 5.641E+01 & 1.294E+01 & 3.418E+00 & 1.934E+00 \\ 
13.6 & 1.618E+03 & 9.707E+02 & 5.351E+02 & 2.057E+02 & 5.958E+01 & 1.344E+01 & 3.485E+00 & 1.955E+00 \\ 
13.7 & 1.763E+03 & 1.052E+03 & 5.770E+02 & 2.200E+02 & 6.294E+01 & 1.396E+01 & 3.554E+00 & 1.976E+00 \\ 
13.8 & 1.921E+03 & 1.140E+03 & 6.221E+02 & 2.353E+02 & 6.650E+01 & 1.450E+01 & 3.625E+00 & 1.998E+00 \\ 
13.9 & 2.094E+03 & 1.235E+03 & 6.709E+02 & 2.516E+02 & 7.028E+01 & 1.506E+01 & 3.697E+00 & 2.020E+00 \\ 
14.0 & 2.282E+03 & 1.339E+03 & 7.237E+02 & 2.692E+02 & 7.429E+01 & 1.565E+01 & 3.772E+00 & 2.042E+00 \\ 
14.1 & 2.487E+03 & 1.451E+03 & 7.807E+02 & 2.881E+02 & 7.854E+01 & 1.627E+01 & 3.848E+00 & 2.065E+00 \\ 
14.2 & 2.712E+03 & 1.574E+03 & 8.423E+02 & 3.083E+02 & 8.305E+01 & 1.691E+01 & 3.927E+00 & 2.089E+00 \\ 
14.3 & 2.957E+03 & 1.706E+03 & 9.089E+02 & 3.300E+02 & 8.784E+01 & 1.758E+01 & 4.007E+00 & 2.112E+00 \\ 
14.4 & 3.224E+03 & 1.850E+03 & 9.810E+02 & 3.532E+02 & 9.292E+01 & 1.828E+01 & 4.090E+00 & 2.137E+00 \\ 
14.5 & 3.516E+03 & 2.007E+03 & 1.059E+03 & 3.782E+02 & 9.832E+01 & 1.901E+01 & 4.175E+00 & 2.161E+00 \\ 
14.6 & 3.834E+03 & 2.177E+03 & 1.143E+03 & 4.050E+02 & 1.040E+02 & 1.978E+01 & 4.262E+00 & 2.186E+00 \\ 
14.7 & 4.182E+03 & 2.362E+03 & 1.234E+03 & 4.338E+02 & 1.101E+02 & 2.058E+01 & 4.352E+00 & 2.212E+00 \\ 
14.8 & 4.561E+03 & 2.562E+03 & 1.333E+03 & 4.647E+02 & 1.166E+02 & 2.142E+01 & 4.444E+00 & 2.238E+00 \\ 
14.9 & 4.976E+03 & 2.780E+03 & 1.439E+03 & 4.979E+02 & 1.234E+02 & 2.229E+01 & 4.539E+00 & 2.264E+00 \\ 
15.0 & 5.428E+03 & 3.017E+03 & 1.555E+03 & 5.336E+02 & 1.307E+02 & 2.321E+01 & 4.636E+00 & 2.291E+00 \\ 
15.1 & 5.923E+03 & 3.274E+03 & 1.679E+03 & 5.719E+02 & 1.385E+02 & 2.416E+01 & 4.736E+00 & 2.319E+00 \\ 
15.2 & 6.462E+03 & 3.554E+03 & 1.814E+03 & 6.130E+02 & 1.467E+02 & 2.516E+01 & 4.839E+00 & 2.347E+00 \\ 
15.3 & 7.052E+03 & 3.858E+03 & 1.961E+03 & 6.572E+02 & 1.554E+02 & 2.621E+01 & 4.944E+00 & 2.376E+00 \\ 
15.4 & 7.696E+03 & 4.189E+03 & 2.119E+03 & 7.047E+02 & 1.647E+02 & 2.730E+01 & 5.053E+00 & 2.405E+00 \\ 
15.5 & 8.399E+03 & 4.548E+03 & 2.290E+03 & 7.557E+02 & 1.746E+02 & 2.845E+01 & 5.164E+00 & 2.434E+00 \\ 
15.6 & 9.168E+03 & 4.939E+03 & 2.475E+03 & 8.106E+02 & 1.851E+02 & 2.965E+01 & 5.279E+00 & 2.464E+00 \\ 
15.7 & 1.001E+04 & 5.364E+03 & 2.676E+03 & 8.696E+02 & 1.963E+02 & 3.090E+01 & 5.396E+00 & 2.495E+00 \\ 
15.8 & 1.092E+04 & 5.826E+03 & 2.893E+03 & 9.330E+02 & 2.082E+02 & 3.221E+01 & 5.517E+00 & 2.526E+00 \\ 
15.9 & 1.193E+04 & 6.328E+03 & 3.129E+03 & 1.001E+03 & 2.208E+02 & 3.359E+01 & 5.642E+00 & 2.558E+00 \\ 
16.0 & 1.302E+04 & 6.874E+03 & 3.384E+03 & 1.074E+03 & 2.343E+02 & 3.503E+01 & 5.770E+00 & 2.591E+00 \\ 
16.1 & 1.422E+04 & 7.468E+03 & 3.660E+03 & 1.153E+03 & 2.486E+02 & 3.653E+01 & 5.901E+00 & 2.624E+00 \\ 
16.2 & 1.553E+04 & 8.115E+03 & 3.959E+03 & 1.238E+03 & 2.638E+02 & 3.811E+01 & 6.036E+00 & 2.658E+00 \\ 
16.3 & 1.695E+04 & 8.818E+03 & 4.283E+03 & 1.329E+03 & 2.800E+02 & 3.976E+01 & 6.176E+00 & 2.692E+00 \\ 
16.4 & 1.852E+04 & 9.583E+03 & 4.634E+03 & 1.427E+03 & 2.972E+02 & 4.150E+01 & 6.319E+00 & 2.727E+00 \\ 
16.5 & 2.022E+04 & 1.042E+04 & 5.014E+03 & 1.533E+03 & 3.156E+02 & 4.331E+01 & 6.466E+00 & 2.763E+00 \\ 
16.6 & 2.209E+04 & 1.132E+04 & 5.427E+03 & 1.647E+03 & 3.351E+02 & 4.521E+01 & 6.617E+00 & 2.799E+00 \\ 
16.7 & 2.413E+04 & 1.231E+04 & 5.873E+03 & 1.769E+03 & 3.559E+02 & 4.720E+01 & 6.773E+00 & 2.836E+00 \\ 
16.8 & 2.636E+04 & 1.338E+04 & 6.357E+03 & 1.900E+03 & 3.781E+02 & 4.929E+01 & 6.933E+00 & 2.874E+00 \\ 
16.9 & 2.880E+04 & 1.455E+04 & 6.882E+03 & 2.042E+03 & 4.017E+02 & 5.147E+01 & 7.098E+00 & 2.913E+00 \\ 
17.0 & 3.147E+04 & 1.582E+04 & 7.451E+03 & 2.195E+03 & 4.268E+02 & 5.377E+01 & 7.267E+00 & 2.952E+00 \\ 
17.1 & 3.438E+04 & 1.720E+04 & 8.068E+03 & 2.359E+03 & 4.536E+02 & 5.617E+01 & 7.442E+00 & 2.992E+00 \\ 
17.2 & 3.757E+04 & 1.871E+04 & 8.737E+03 & 2.536E+03 & 4.821E+02 & 5.869E+01 & 7.621E+00 & 3.033E+00 \\ 
17.3 & 4.105E+04 & 2.034E+04 & 9.462E+03 & 2.726E+03 & 5.125E+02 & 6.134E+01 & 7.806E+00 & 3.074E+00 \\ 
17.4 & 4.487E+04 & 2.213E+04 & 1.025E+04 & 2.931E+03 & 5.449E+02 & 6.411E+01 & 7.996E+00 & 3.116E+00 \\ 
17.5 & 4.904E+04 & 2.407E+04 & 1.110E+04 & 3.152E+03 & 5.795E+02 & 6.702E+01 & 8.192E+00 & 3.160E+00 \\ 
17.6 & 5.360E+04 & 2.619E+04 & 1.203E+04 & 3.390E+03 & 6.163E+02 & 7.007E+01 & 8.393E+00 & 3.204E+00 \\ 
17.7 & 5.858E+04 & 2.849E+04 & 1.303E+04 & 3.647E+03 & 6.556E+02 & 7.327E+01 & 8.601E+00 & 3.248E+00 \\ 
17.8 & 6.404E+04 & 3.100E+04 & 1.412E+04 & 3.923E+03 & 6.975E+02 & 7.663E+01 & 8.814E+00 & 3.294E+00 \\ 
17.9 & 7.001E+04 & 3.374E+04 & 1.530E+04 & 4.221E+03 & 7.421E+02 & 8.015E+01 & 9.034E+00 & 3.341E+00 \\ 
18.0 & 7.654E+04 & 3.671E+04 & 1.658E+04 & 4.541E+03 & 7.897E+02 & 8.385E+01 & 9.261E+00 & 3.388E+00 \\ 
18.1 & 8.368E+04 & 3.996E+04 & 1.797E+04 & 4.887E+03 & 8.405E+02 & 8.774E+01 & 9.494E+00 & 3.437E+00 \\ 
18.2 & 9.149E+04 & 4.349E+04 & 1.948E+04 & 5.260E+03 & 8.947E+02 & 9.182E+01 & 9.734E+00 & 3.486E+00 \\ 
18.3 & 1.000E+05 & 4.734E+04 & 2.112E+04 & 5.662E+03 & 9.525E+02 & 9.610E+01 & 9.981E+00 & 3.537E+00 \\ 
18.4 & 1.094E+05 & 5.153E+04 & 2.290E+04 & 6.095E+03 & 1.014E+03 & 1.006E+02 & 1.024E+01 & 3.588E+00 \\ 
18.5 & 1.196E+05 & 5.610E+04 & 2.483E+04 & 6.563E+03 & 1.080E+03 & 1.053E+02 & 1.050E+01 & 3.640E+00 \\ 
18.6 & 1.308E+05 & 6.108E+04 & 2.692E+04 & 7.066E+03 & 1.150E+03 & 1.103E+02 & 1.077E+01 & 3.694E+00 \\ 
18.7 & 1.431E+05 & 6.651E+04 & 2.920E+04 & 7.610E+03 & 1.225E+03 & 1.155E+02 & 1.105E+01 & 3.748E+00 \\ 
18.8 & 1.565E+05 & 7.242E+04 & 3.167E+04 & 8.196E+03 & 1.305E+03 & 1.210E+02 & 1.133E+01 & 3.804E+00 \\ 
18.9 & 1.712E+05 & 7.887E+04 & 3.435E+04 & 8.828E+03 & 1.391E+03 & 1.267E+02 & 1.163E+01 & 3.860E+00 \\ 
19.0 & 1.873E+05 & 8.589E+04 & 3.726E+04 & 9.510E+03 & 1.482E+03 & 1.328E+02 & 1.194E+01 & 3.918E+00 \\ 
19.1 & 2.048E+05 & 9.355E+04 & 4.042E+04 & 1.025E+04 & 1.580E+03 & 1.391E+02 & 1.225E+01 & 3.977E+00 \\ 
19.2 & 2.241E+05 & 1.019E+05 & 4.385E+04 & 1.104E+04 & 1.684E+03 & 1.458E+02 & 1.257E+01 & 4.037E+00 \\ 
19.3 & 2.452E+05 & 1.110E+05 & 4.758E+04 & 1.190E+04 & 1.795E+03 & 1.528E+02 & 1.291E+01 & 4.099E+00 \\ 
19.4 & 2.682E+05 & 1.209E+05 & 5.163E+04 & 1.282E+04 & 1.914E+03 & 1.602E+02 & 1.325E+01 & 4.161E+00 \\ 
19.5 & 2.935E+05 & 1.317E+05 & 5.603E+04 & 1.382E+04 & 2.041E+03 & 1.679E+02 & 1.361E+01 & 4.225E+00 \\ 
19.6 & 3.211E+05 & 1.435E+05 & 6.081E+04 & 1.490E+04 & 2.176E+03 & 1.761E+02 & 1.397E+01 & 4.290E+00 \\ 
19.7 & 3.514E+05 & 1.564E+05 & 6.600E+04 & 1.606E+04 & 2.321E+03 & 1.847E+02 & 1.435E+01 & 4.356E+00 \\ 
19.8 & 3.845E+05 & 1.704E+05 & 7.164E+04 & 1.731E+04 & 2.476E+03 & 1.937E+02 & 1.474E+01 & 4.424E+00 \\ 
19.9 & 4.208E+05 & 1.857E+05 & 7.777E+04 & 1.867E+04 & 2.642E+03 & 2.032E+02 & 1.514E+01 & 4.493E+00 \\ 
20.0 & 4.606E+05 & 2.024E+05 & 8.444E+04 & 2.013E+04 & 2.819E+03 & 2.132E+02 & 1.555E+01 & 4.564E+00 \\ 
\end{longtable*}
\end{center}
\newpage

\begin{deluxetable}{rrrrrrrrrr}
\centering
\tabletypesize{\footnotesize}
\tablecaption{[OIII] Line flux ratios vs kinetic temperature $T_U$ (see Fig. 8)\label{table_2}}
\tablehead{
\colhead{$T_U$ (\degr K)} & \colhead{$\kappa=2$} & \colhead{$\kappa=3$} & \colhead{$\kappa=4$} & 
\colhead{$\kappa=6$} & \colhead{$\kappa=10$} & \colhead{$\kappa=20$} & \colhead{$\kappa=50$} & \colhead{$\kappa=100$} & \colhead{Maxwell} \\%[-1ex]
}
\startdata
5000 & 29.7012 & 48.1983 & 72.5995 & 138.179 & 322.942 & 914.582 & 2365.64 & 3551.70 & 5678.96 \\ 
5500 & 29.4789 & 46.8930 & 69.0278 & 125.368 & 270.262 & 674.143 & 1513.02 & 2121.70 & 3117.32 \\ 
6000 & 29.2609 & 45.6613 & 65.7814 & 114.480 & 229.899 & 515.460 & 1032.33 & 1372.76 & 1891.11 \\ 
6500 & 29.0469 & 44.4974 & 62.8208 & 105.148 & 198.350 & 406.285 & 741.598 & 945.508 & 1238.84 \\ 
7000 & 28.8369 & 43.3962 & 60.1122 & 97.0852 & 173.257 & 328.495 & 555.442 & 684.565 & 861.973 \\ 
7500 & 28.6308 & 42.3530 & 57.6269 & 90.0701 & 152.988 & 271.388 & 430.514 & 516.106 & 629.358 \\ 
8000 & 28.4285 & 41.3637 & 55.3402 & 83.9261 & 136.389 & 228.376 & 343.322 & 402.263 & 477.847 \\ 
8500 & 28.2299 & 40.4242 & 53.2305 & 78.5127 & 122.629 & 195.251 & 280.418 & 322.330 & 374.680 \\ 
9000 & 28.0348 & 39.5313 & 51.2794 & 73.7162 & 111.094 & 169.241 & 233.734 & 264.362 & 301.772 \\ 
9500 & 27.8433 & 38.6815 & 49.4707 & 69.4446 & 101.329 & 148.467 & 198.235 & 221.147 & 248.601 \\ 
10000 & 27.6551 & 37.8721 & 47.7904 & 65.6223 & 92.9877 & 131.622 & 170.664 & 188.154 & 208.767 \\ 
10500 & 27.4702 & 37.1003 & 46.2259 & 62.1868 & 85.8035 & 117.779 & 148.851 & 162.440 & 178.225 \\ 
11000 & 27.2886 & 36.3637 & 44.7665 & 59.0863 & 79.5699 & 106.266 & 131.311 & 142.033 & 154.330 \\ 
11500 & 27.1102 & 35.6602 & 43.4024 & 56.2774 & 74.1239 & 96.5858 & 117.002 & 125.577 & 135.302 \\ 
12000 & 26.9348 & 34.9875 & 42.1252 & 53.7234 & 69.3362 & 88.3683 & 105.179 & 112.119 & 119.912 \\ 
12500 & 26.7623 & 34.3438 & 40.9273 & 51.3936 & 65.1029 & 81.3305 & 95.2955 & 100.973 & 107.292 \\ 
13000 & 26.5928 & 33.7274 & 39.8019 & 49.2613 & 61.3397 & 75.2547 & 86.9487 & 91.6369 & 96.8137 \\ 
13500 & 26.4262 & 33.1366 & 38.7429 & 47.3043 & 57.9780 & 69.9709 & 79.8335 & 83.7378 & 88.0182 \\ 
14000 & 26.2623 & 32.5699 & 37.7450 & 45.5029 & 54.9612 & 65.3451 & 73.7166 & 76.9927 & 80.5612 \\ 
14500 & 26.1012 & 32.0260 & 36.8034 & 43.8406 & 52.2424 & 61.2705 & 68.4173 & 71.1849 & 74.1819 \\ 
15000 & 25.9426 & 31.5035 & 35.9135 & 42.3027 & 49.7824 & 57.6609 & 63.7940 & 66.1462 & 68.6799 \\ 
15500 & 25.7867 & 31.0013 & 35.0715 & 40.8766 & 47.5483 & 54.4468 & 59.7342 & 61.7444 & 63.8991 \\ 
16000 & 25.6333 & 30.5183 & 34.2739 & 39.5513 & 45.5125 & 51.5708 & 56.1483 & 57.8745 & 59.7166 \\ 
16500 & 25.4823 & 30.0534 & 33.5173 & 38.3170 & 43.6512 & 48.9858 & 52.9635 & 54.4524 & 56.0347 \\ 
17000 & 25.3338 & 29.6057 & 32.7988 & 37.1652 & 41.9444 & 46.6528 & 50.1207 & 51.4099 & 52.7748 \\ 
17500 & 25.1876 & 29.1742 & 32.1158 & 36.0884 & 40.3747 & 44.5389 & 47.5714 & 48.6915 & 49.8733 \\ 
18000 & 25.0437 & 28.7582 & 31.4658 & 35.0798 & 38.9272 & 42.6167 & 45.2751 & 46.2514 & 47.2781 \\ 
18500 & 24.9020 & 28.3569 & 30.8466 & 34.1336 & 37.5891 & 40.8628 & 43.1986 & 44.0517 & 44.9463 \\ 
19000 & 24.7625 & 27.9695 & 30.2562 & 33.2444 & 36.3492 & 39.2574 & 41.3137 & 42.0609 & 42.8424 \\ 
19500 & 24.6252 & 27.5953 & 29.6926 & 32.4074 & 35.1975 & 37.7835 & 39.5966 & 40.2524 & 40.9365 \\ 
20000 & 24.4900 & 27.2337 & 29.1543 & 31.6185 & 34.1256 & 36.4266 & 38.0273 & 38.6038 & 39.2038 \\ 
\enddata
\tablecomments{(1) Maxwell figures computed using Lennon \& Burke (1994) $\Omega$ data.}
\end{deluxetable}

\end{document}